\documentclass[aip,jcp,preprint,longbibliography,floatfix]{revtex4-2}

\usepackage{graphicx}
\usepackage{amsmath, amssymb}% Include figure files
\usepackage{mathtools}
\usepackage{hyperref}
\usepackage{color}
\usepackage{ulem}
\def\ptl{\partial}
\def\ie{\textit{i.e.}, }
\def\eg{\textit{e.g.}, }
\def\kB{k_\mathrm{B}}
\def\drom{\mathrm{d}}
\def\eqref#1{(\ref{#1})}
\def\angb#1{\left<#1\right>}

\def\tfk{t_\mathrm{fk}}
\def\tslip{t_\mathrm{slip}}
\def\tsys{t_\mathrm{sys}}
\def\Fw{F_\mathrm{w}}
\def\uslip{u_\mathrm{slip}}
\def\tauw{\tau_\mathrm{w}}
\def\uw{u_\mathrm{w}}
\def\lammax{\lambda_\mathrm{max}}
\def\lamshort{\lambda_\mathrm{short}}
\def\lammid{\lambda_\mathrm{mid}}
\def\lamtail{\lambda_{\infty}}
\def\lamnemd{\lambda_\mathrm{NEMD}}
\def\Lammid{\Lambda_\mathrm{mid}}
\def\Lamshort{\Lambda_\mathrm{short}}
\def\epsfs{\varepsilon^\mathrm{fs}}
\def\epsff{\varepsilon^\mathrm{ff}}
\def\sigfs{\sigma^\mathrm{fs}}

\newcommand{\pd}[2]{\frac{\partial #1}{\partial #2}}
\newcommand{\dd}[2]{\frac{\mathrm{d}\, #1}{\mathrm{d}\, #2}}
\newcommand{\pdd}[2]{\frac{\partial^2 #1}{\partial {#2}^2}}

\newcommand{\bracket}[1]{\left< #1 \right>}
\newcommand{\braced}[1]{\left( #1 \right)}

\newcommand{\fl}[1]{\tilde{#1}}
\newcommand{\rd}{\,\mathrm{d}}

\newcommand{\osaka}{Department of Mechanical Engineering, Osaka University, 2-1 Yamadaoka, Suita 565-0871, Japan}
\newcommand{\tuswater}{Water Frontier Research Center (WaTUS),
Research Institute for Science \& Technology,
Tokyo University of Science,
1-3 Kagurazaka, Shinjuku-ku, Tokyo, 162-8601, Japan}
\newcommand{\osakap}{Department of Mechanical Engineering, 
Osaka Metropolitan University, 3-3-138 Sugimoto, Sumiyoshi, Osaka 558-8585, Japan}
\newcommand{\ilm}{Univ Lyon, Univ Claude Bernard Lyon 1, CNRS, Institut Lumi\`ere Mati\`ere, F-69622, VILLEURBANNE, France}
\begin{document}
%\preprint{}

\title{%\add{%
Equilibrium molecular dynamics evaluation of the solid-liquid friction coefficient: role of timescales
%}
}

\author{Haruki Oga}%
\email{haruki@nnfm.mech.eng.osaka-u.ac.jp}
\affiliation{\osaka}
\author{Takeshi Omori}%
\email{t.omori@omu.ac.jp}
\affiliation{\osakap}
\author{Laurent Joly}%
\email{laurent.joly@univ-lyon1.fr}
\affiliation{\ilm}
\author{Yasutaka Yamaguchi}
\email{yamaguchi@mech.eng.osaka-u.ac.jp}
%\homepage{http://www-nnfm.mech.eng.osaka-u.ac.jp/~yamaguchi/}
\affiliation{\osaka}
\affiliation{\tuswater}
\date{\today}

\begin{abstract}
Solid-liquid friction plays a key role in nanofluidic systems. Yet, despite decades of method development to quantify solid-liquid friction using molecular dynamics (MD) simulations, an accurate and widely applicable method is still missing. Here,
we propose a method to quantify the solid-liquid friction coefficient (FC) 
from equilibrium MD simulations of a liquid confined between parallel solid walls. 
In this method, the FC is evaluated by fitting the Green-Kubo (GK) integral of the S-L shear force autocorrelation for the range of time scales where the GK integral slowly decays with time. 
The fitting function was derived 
based on
the analytical solution considering 
the hydrodynamic equations in our previous work [H. Oga et al., Phys. Rev. Research \textbf{3}, L032019 (2021)], assuming that the timescales related to the 
friction kernel and to the bulk viscous dissipation can be separated. 
By comparing the results with those of other equilibrium MD-based methods and those of non-equilibrium MD for a Lennard-Jones liquid between flat crystalline walls with different wettability, we show that the FC is extracted with excellent accuracy by the present method, even in wettability regimes where other methods become innacurate. We then show that the method is also applicable to grooved solid walls, for which the GK integral displays a complex behavior at short times. Overall, the present method extracts efficiently the FC for various systems, with easy implementation and low computational cost. 
\end{abstract}

\maketitle %\maketitle must follow title, authors, abstract and \pacs
\section{Introduction}
\label{sec:intro}
Nanofluidics describes the motion of fluids confined at the nanoscale, 
and plays an important role in various fields such as nanotechnology, biology, and energy conversion.\cite{Eijkel2005,Sparreboom2009,Bocquet2010,Schoch2008,Sparreboom2010,Striolo2016} 
Solid-liquid (S-L) slip particularly affects the fluid transport,
and Navier proposed the following boundary condition (BC) as the slip model:\cite{Navier1823} 
\begin{equation}
   \tau_{\text{w}}=\lambda_0 u_{\text{slip}},
   \label{eq:navier}
\end{equation}
where $\tau_{\text{w}}$ is the S-L friction force per area,
$\lambda_0$ is the S-L friction coefficient (FC) and $u_{\text{slip}}$ is the slip velocity. Equation~\eqref{eq:navier} is called the Navier BC. 
If Newton's law of viscosity is applied at the S-L interface, the shear force per area, \ie shear stress, is given by 
\begin{equation}\label{eq:newtonvisc}
    \tauw = \eta \left. \pd{u}{z} \right|_\text{interface},
\end{equation}
where $\eta$ and $u$ are the liquid viscosity and the velocity parallel to the S-L interface as a function of the normal position $z$.
Hence, the Navier BC in Eq.~\eqref{eq:navier} can be written in another form as
\begin{equation}
    \frac{u_{\text{slip}}}{b} = \left. \pd{u}{z}\right|_{\text{interface}},
\end{equation}
where $b$ is called the slip length and is defined by
\begin{equation}\label{eq:def_b}
 b  =\frac{\eta}{\lambda_0}.
\end{equation}
%
%\red{
The slip length depends on the S-L combination; \eg for water on a graphene or carbon nanotube surface, $b$ was theoretically and experimentally estimated to be about several tens of nanometers.\cite{Falk2010,Keerthi2021,Chen2021}  
%}
%
\par
Molecular dynamics (MD) is a powerful tool to evaluate the FC or the slip length, and to explore the mechanisms underlying S-L friction.\cite{Huang2008,Ogawa2019,Thompson1990,Thompson1997,Barrat1999,Cieplak2001,Falk2010,Kannam2013,Bhatia2013,Tocci2014,Guo2016,Nakaoka2017,Ewen2019} 
The FC can be directly calculated by using non-equilibrium MD (NEMD) simulations of Poiseuille or Couette flows; 
however, this requires a high shear rate, typically above $10^9$~s$^{-1}$, to reduce the statistical error due to thermal fluctuation. In such range, the FC $\lambda_{0}$ can depend on the shear rate, \ie the shear force $\tauw$ is not proportional to the slip velocity $\uslip$ as in Eq.~\eqref{eq:navier}.\cite{Thompson1997,Kannam2011,KumarKannam2012} 
In addition, the S-L interface position must be defined to strictly determine the slip velocity $\uslip$ or the slip length $b$, which is not trivial.\cite{Herrero2019} 
\par
On the other hand, several methods for calculating the FC
from equilibrium MD (EMD) with a zero shear rate have been proposed.\cite{Bocquet1994,Petravic2007,Hansen2011,KumarKannam2012,Bocquet2013,Huang2014,Sam2018,Oga2019,Varghese2021,Nakano2020,Sokhan2008, Hadjiconstantinou2022} 
In their pioneering work, Bocquet and Barrat (BB) \cite{Bocquet1994,Bocquet2013} introduced a Green-Kubo (GK) integral defined by
\begin{equation}
%  \lambda_0 = \Lambda (t_\text{plateau}), \text{ with }\ 
  \Lambda (t) \equiv \frac{1}{S k_{\text{B}} T} \int_0^t
  \langle \Fw(t) \Fw(0) \rangle \drom t ,
  \label{eq:def_GKint}
\end{equation}
where $S$, $\kB$, $T$ and $\angb{ \Fw(t)\Fw(0)}$
denote the surface area, Boltzmann constant,
absolute temperature and equilibrium autocorrelation function of the instantaneous shear force $\Fw$ on the solid as a function of time $t$, respectively, for the calculation of the FC. 
%\red{
In bulk systems, GK integrals are a standard tool to calculate transport properties,
%} 
%Indeed, using the integral of the autocorrelation function, \ie the GK integral, is a key feature of the Green-Kubo relation used to calculate for bulk transport properties, 
\eg the viscosity of a fluid can be obtained from the convergence value autocorrelation of the off-diagonal stress component.\cite{Evans2008book} 
However, in contrast to bulk GK integrals, which show a simple behavior of monotonically increasing with time $t$ and converging to a certain value for $t\rightarrow \infty$, $\Lambda(t)$ typically increases for a short time, and decreases after taking a maximum, 
%\red{
which we will call the intermediate plateau value,
%} 
%called the intermediate plateau value, 
and usually converges to a non-zero final plateau value for $t \rightarrow \infty$. This behavior is 
%\red{
often referred to as
%}
%called 
the plateau problem.\cite{Oga2019, Espanol2019, Merabia2012} 
%
%\rem{
%Bocquet and Barrat proposed that the S-L friction coefficient in the Navier BC in Eq.~\eqref{eq:navier} can be approximated by the intermediate plateau value.
%}
%
%\begin{equation}
%  \lambda_0 \approx \Lambda (t_\text{plateau}).
%  \label{eq:GK_BB}
%\end{equation}
%
%Especially 
When the friction coefficient is small, \ie the slip length is large, the 
%\red{
intermediate
%} 
plateau region of the GK integral $\Lambda(t)$ is clearly observed, %and the plateau value can be easily determined, 
and in such cases, it indeed gives a good estimate of the corresponding NEMD result.\cite{Oga2019} 
However, for larger FC, $\Lambda(t)$ decays faster with time after taking a maximum, thus, the 
%\red{
intermediate
%} 
plateau region is not apparent, 
%\red{
and the corresponding value of $\Lambda (t)$
%} 
%and  $\Lambda (t_\text{plateau})$ 
is not well-defined.\cite{Oga2019, Espanol2019, Merabia2012} 
Even in such case, the FC is often estimated from the maximum value of the GK integral as \cite{Nakaoka2017b,Oga2019,Nakano2020} 
\begin{equation}\label{eq:lambda0=Lambdamax}
    \lambda_{0} \approx \max\left[\Lambda(t)\right],
\end{equation}
although it does not necessarily give a proper estimate.\cite{Oga2019} 
\par
Recently, the authors derived %a solution 
%\blue{
an analytical expression
%}
of the GK integral $\Lambda(t)$
by explicitly modelling the liquid motion described by the Stokes equation.
The %analytical expression of the GK integral 
expression 
$\Lambda(t)$ includes a non-Markovian effect %called 
%\red{
quantified by
%}
the friction kernel $\lambda(t)$,~\cite{Schulz2020,Omori2019} 
with which the friction force per area $\tauw(t)$ at time $t$ is expressed by 
\begin{equation}\label{eq:friction-kernel}
    \tauw(t) = \int_{-\infty}^{t} \lambda(t-s) u_\text{slip}(s) \drom s, 
\end{equation}
including the hysteresis dependence of the slip velocity.
For a steady flow with a constant slip velocity $\uslip$, 
the friction coefficient $\lambda_{0}$ in the Navier BC is related to this friction kernel $\lambda(t)$ by
\begin{equation}\label{eq:def_lambda0}
    \lambda_0 = \int_0^\infty \lambda(t)\drom t. 
\end{equation}
%
%\red{
For simple liquids at room temperature,
%} 
%although 
the friction kernel $\lambda(t)$ typically decays within a short timescale, which we denote by $\tfk$, around several picoseconds.
Considering this feature, the friction kernel is often modeled by the following Maxwell-type expression:\cite{Omori2019, Oga2021,Varghese2021} 
\begin{equation}\label{eq:friction-Maxwell}
    \lambda(t) = \frac{\lambda_0}{\tfk} e^{-\frac{t}{\tfk}} .
\end{equation}
%
%\red{
Note that taking the limit $\tfk \rightarrow 0$ corresponds to a Markovian FC without hysteresis effect.
%} 
It has been shown that the Maxwell model in Eq.~\eqref{eq:friction-Maxwell} approximates well the friction for various kinds of liquids, \eg Lennard-Jones (LJ) liquids or water, on various solid surfaces,\cite{Omori2019, Oga2021,Varghese2021} 
and that $\Lambda(t)$ typically increases %up 
from zero %up to this timescale 
and takes the maximum 
%\blue{
around $\tfk$,
%}, 
whilst for specific cases such as supercooled water, the simple Maxwell-type kernel is not sufficient to express $\lambda(t)$.\cite{Oga2021} 
%
%\red{
Related to this point, 
Hansen et al. \cite{Hansen2011,Varghese2021} proposed that the friction kernel 
$\lambda(t)$ is calculated from Eq.~\eqref{eq:friction-kernel}
by measuring the fluctuations of the S-L friction force and the slip velocity, 
assuming a Maxwell-type friction kernel $\lambda(t)$. % to be the Maxwell type in Eq.~\eqref{eq:friction-Maxwell}.
%}
In addition, 
\citet{Nakano2020} derived an analytical expression of the GK integral $\Lambda(t)$ based on linearized fluctuating hydrodynamics (LFH) by assuming timescale separation, % as in this study,
and they proposed a measurement method of $\lambda_0$ by fitting the GK integral.
%\red{Including these mention about limitation of the existing methods + add discussion part of res-discussion}
\par
%\red{
Overall, the understanding of the solid-liquid FC and of the related GK integral has progressed significantly during the last years; however, a practical method to extract the FC based on EMD simulations with sufficient accuracy as well as with a wide applicability is still missing.
%}
In this study, we propose a new measurement method of the FC
based on time separation %of the solution of 
in
the GK integral $\Lambda(t)$, and %\red{
we show the %
advantages
%interest 
of this method by comparing its results
%} 
%examined the validity through the comparison 
with NEMD results as well as existing EMD-based results.

\section{Theory}
\label{sec:theory}
%\input{2-theory.tex}
%
%\red{detail should be to Appendix:} 
Let us consider a system where a liquid is confined between two fixed planar solid walls.
The authors derived a theoretical expression of the equilibrium autocorrelation function of the S-L friction force $C_{\Fw}(t) \equiv \angb{\Fw(t) \Fw(0)}$ for that system 
-- which is the differential of the GK integral $\Lambda(t)$ as shown in Eq.~\eqref{eq:def_GKint} -- 
by coupling the Stokes equation for the liquid motion and a Langevin equation for the motion of one of the  walls in the direction parallel to the interface, using the Fourier-Laplace (FL) transform as \cite{Oga2021}
\begin{equation}\label{eq:GK-full}
    \frac{%
        \fl{C}_{\Fw}(\omega)
    }
    {%
        S\kB T
    }
    =
	\frac{%
    	\fl{\lambda} \eta\zeta
    	\left[
        	\eta\zeta \sinh(\zeta h) + 
        	\fl{\lambda} \cosh(\zeta h)
    	\right]
	}
	{%
    	(\fl{\lambda}^2 + \eta^2 \zeta^2)\sinh(\zeta h) + 
    	2 \fl{\lambda}\eta\zeta \cosh(\zeta h)
	},
\end{equation}
where the FL transform denoted by tilde is defined by
\begin{equation}\label{eq:def_FL}
    \tilde f (\omega) \equiv \int_0^\infty f(t)e^{-i\omega t} \text d t,
\end{equation}
and $\zeta$ is given by
\begin{equation}
    \zeta = \sqrt{\frac{i\rho\omega}{\eta}},
    \label{eq:def_zeta}
\end{equation}
with $\rho$, $h$ and $\lambda$ being the fluid density, the distance between the top and bottom S-L interfaces, and the friction kernel, respectively (see Appendix~\ref{appsubsec:Lang-Stokes}).
\par
However, Eq.~\eqref{eq:GK-full} in the FL form 
%as well as including $\zeta$ in Eq.~\eqref{eq:def_zeta} 
is rather complex and the solution does not give clear outlook 
of the physical aspects of S-L friction. More practically, calculating the FC 
$\lambda_{0}$ %in the Navier BC in Eq.~\eqref{eq:navier}
directly from Eq.~\eqref{eq:GK-full} is not trivial.
Regarding this point, in our previous study,\cite{Oga2021} we proposed to use the convergence value $\Lambda(\infty)$ 
as one possible method to obtain $\lambda_{0}$ (see Appendix~\ref{appsubsec:Lambda_infty}), where $\Lambda(\infty)$ was related to $\lambda_{0}$ as:
%
%Eq.~\eqref{eq:GK-full-conv} results in
%
\begin{align}\label{eq:GK-full-conv2}
\Lambda(\infty) \equiv 
\lim_{t\rightarrow\infty} \Lambda(t)
	= \frac{\lambda_0}{\frac{h}{\eta}\lambda_0+2}
	= \frac{\lambda_0}{\frac{h}{b}+2}.
\end{align}
%
%where Eq.~\eqref{eq:def_b} is used for the final equality.
%
Thus, $\lambda_0$ can be evaluated  by
\begin{equation}
\label{eq:Lambda_conv}
    \lambda_0 = \frac{2\Lambda(\infty)}{1-\frac{h}{\eta}\Lambda(\infty)}.
\end{equation}
Equation~\eqref{eq:Lambda_conv} indicates that the convergence value of the GK integral $\Lambda(t)$ depends on the liquid height $h$, and this partly has given an answer to the long-standing issue of the plateau problem of the GK integral.\cite{Oga2021} 
From Eq.~\eqref{eq:GK-full-conv2}, it is also clear that a semi-infinite system with $h \rightarrow \infty$ results in $\Lambda(\infty) \rightarrow 0$, 
%\red{
and that the final plateau value $\Lambda(\infty)$ and $\lambda_{0}$ are obviously different. In particular, from Eq.~\eqref{eq:GK-full-conv2} one can see that when $b/h \rightarrow \infty$, $\Lambda(\infty) \rightarrow \lambda_0/2$, so that even in that limit the final plateau does not identify with the friction coefficient. In practice however, for very large slip lengths the final plateau appears after a very long time, and only the intermediate plateau can be observed in the simulations, whose value is indeed $\lambda_0$.\cite{Oga2019}
%}
%
\par
Although Eq.~\eqref{eq:Lambda_conv} is simple and insightful, using the convergence value $\Lambda(\infty)$ of the long-range time integration of the correlation function in EMD systems 
%\red{
is difficult in practice: for small friction coefficients, the time needed to reach the final plateau becomes very large, and so does the statistical error on $\Lambda(\infty)$; for large friction coefficients, the denominator in Eq.~\eqref{eq:Lambda_conv} becomes close to zero, leading also to large uncertainties on $\lambda_0$.
%}  
%in practice is disadvantageous because the statistical error becomes inevitably large as the time $t$ increases for non-zero $\Lambda(\infty)$.\cite{Oga2021} 
In addition, the viscosity $\eta$ must be additionally provided,  typically calculated using another system, and the liquid
%\red{
hydrodynamic
%} 
height $h$ must also be determined. %, \ie $h$ is not exactly the same as the liquid height of the MD system but should be determined by another simulation. 
Regarding the latter point, it was indicated that the 
%\red{
hydrodynamic position of the S-L interface (where the Navier BC applies)
%} 
was approximately one liquid particle diameter %of the liquid particle 
outward the wall surface for the case of a Lennard-Jones liquid on a flat surface,\cite{Herrero2019, Omori2019,Oga2021} whilst determining the interface position for complex surfaces is not trivial.
%\red{Furthermore, large FC $\lambda_{0}$ cannot be properly evaluated by Eq.~\eqref{eq:Lambda_conv} as understood from its function form and Eq.~\eqref{eq:GK-full-conv2}.}
%
\par
In this study, we propose a new method to measure the FC by simplifying Eq.~\eqref{eq:GK-full} through considering the timescales of the friction kernel.
%\red{
At first, we consider the limit $\tfk \rightarrow +0$, \ie a Markovian Navier BC. %in Eq.~\eqref{eq:navier}. 
Then, the FL transform $\tilde \lambda$ of Eq.~\eqref{eq:friction-Maxwell} writes
\begin{equation}
  \lim_{\tfk \rightarrow +0} \fl{\lambda}(\omega)
  =
  \lambda_{0} .
\end{equation}
%At first, we assume the Maxwell-type FC in Eq.~\eqref{eq:friction-Maxwell} with the limit $\tfk \rightarrow 0$, \ie Markovian Navier B.C. in Eq.~\eqref{eq:navier}. 
%Then the FL tranform $\tilde \lambda$ of Eq~\eqref{eq:friction-Maxwell} writes
%%
%\begin{equation}
%  \fl{\lambda}(\omega) = \lim_{\tfk \rightarrow 0} 
%  \frac{\lambda_{0}}{1+i\omega \tfk}
%  =
%  \lambda_{0}
%\end{equation}
%}
%
In addition, we consider the limit $h\rightarrow \infty$ in 
Eq.~\eqref{eq:GK-full}. Then, under these two limits, it follows from Eq.~\eqref{eq:GK-full} that 
\begin{equation}\label{eq:GK-full-twolimits}
%  \fl{\Lambda}_\mathrm{mid}(\omega)
%  \equiv
    \lim_{\tfk \rightarrow +0}
    \lim_{h \rightarrow \infty}
    \frac{%
        \fl{C}_{\Fw}(\omega)
    }
    {%
        S\kB T
    }
    =
	\frac{%
    	\lambda_{0} \eta\zeta
    % 	\left[
    %     	\eta\zeta %\sinh(\zeta h) 
    %     	+ 
    %     	\lambda_{0} 
    %     	%\cosh(\zeta h)
    % 	\right]
	}
	{%
    	\eta \zeta + \lambda_{0}
    	%^2 + \eta^2 \zeta^2%\sinh(\zeta h) 
    	%+ 
    	%2 \lambda_{0}\eta\zeta 
    	%\cosh(\zeta h)
	}.
\end{equation}
The inverse FL transform of the 
RHS in Eq.~\eqref{eq:GK-full-twolimits}
is analytically obtained as
%,  and $\lambda(t)=\lambda_0 \delta(t)$ on Eq.~\eqref{eq:GK-full}:
%
\begin{equation}
    \lim_{\tfk \rightarrow +0}
    \lim_{h \rightarrow \infty}
    \frac{%
        C_{\Fw}(\omega)
    }
    {%
        S\kB T
    }
    =\frac{\lambda_{0}}{\tslip}
    \exp{\braced{\frac{t}{\tslip}}}
    \text{erfc}
    \braced{\sqrt{\frac{t}{\tslip}}}
    -
    \frac{\lambda_{0}}{\sqrt{\pi \tslip t}}
    ,
\end{equation}
%\red{
introducing a second timescale $\tslip$, given by
%\red{introducing a third timescale $\tslip$, -- other than the kernel timescale $\tfk$ and the sytem timescale $\tsys$ related to the system size mentioned below --, given by
\begin{equation}\label{eq:def_tslip}
\tslip=\frac{\rho b^2}{\eta}=\frac{\rho\eta}{\lambda_{0}^{2}} , 
\end{equation}
which can be interpreted as the time of diffusion of the momentum in the liquid over the slip length.
%}
%
%\red{As mentioned above, taking the limit $\tfk \rightarrow 0$ corresponds to a Markovian FC without hysteresis effect.}

Then, the limit of GK integral $\Lammid(t)$ results in 
\begin{equation}\label{eq:GK-mid}
    \Lambda_\text{mid}(t) 
    \equiv
    \lim_{\tfk \rightarrow +0}
    \lim_{h \rightarrow \infty}
    \int_0^t
    \frac{%
        C_{\Fw} (t)
    }
    {%
        S\kB T
    }
    dt
    =
%   \!\!\!\!\!\!
%    \lim_{\scriptstyle{\begin{matrix}
%    \scriptstyle h\rightarrow\infty,\\
%    \scriptstyle \lambda(t)\rightarrow\lambda_0\delta(t)
%   \end{matrix}}}
%    \!\!\!\!\!\!
%    = 
    \lambda_0 \exp\braced{\frac{t}{\tslip}} \text{erfc} \braced{\sqrt{\frac{t}{\tslip}}}.
\end{equation}
%
% where $\tslip$ given by% \red{mention about BB2013 shortly}
% %
% \begin{equation}\label{eq:def_tslip}
% \tslip=\frac{\rho b^2}{\eta}=\frac{\rho\eta}{\lambda_{0}^{2}}
% \end{equation}
% %
% is the third timescale included in the solution $\Lammid(t)$, which corresponds to the diffusion timescale of the streaming velocity to the slip length $b$.
%
Equation~\eqref{eq:GK-mid} is the key to obtain the FC from the GK integral $\Lambda(t)$ in this study. 
%\red{
Note that \citet{Bocquet2013} showed an expression similar to Eq.~\eqref{eq:GK-mid} %including the complementary error function erfc 
for a semi-infinite system, \ie $h\rightarrow \infty$.
%}
%We fit the used this function form $\Lambda^\text{mid}(t)$ for the fitting 
%
\par
We see the meaning of the solution including a timescale $\tslip$ with respect to the limits related to the other two timescales: the Markovian limit with $\tfk \rightarrow +0$
%in Eq.~\eqref{eq:friction-Maxwell}
and infinite height $h \rightarrow \infty$.
The first is supposed to be applicable if the timescale of interest %being observed 
is sufficiently longer than the decay timescale
of $\tfk$. For the second limit $h\rightarrow \infty$, we consider the diffusion timescale of the %streaming 
velocity information generated on one wall to the opposite wall, given by
\begin{equation}\label{eq:def_tsys}
    \tsys \equiv \frac{\rho h^2}{\eta}.
\end{equation}
This has the same form as $\tslip$ in Eq.~\eqref{eq:def_tslip}, with substituting the slip length $b$ by  $h$, corresponding to a system timescale defined by the finite liquid height $h$.
%
%Thus, the second limit $h\rightarrow \infty$ is supposed to be applicable for a timescale shorter than $\tsys$.
%
%the two limits $\tilde\lambda \rightarrow \lambda_0$ and $h\rightarrow \infty$ in Eq.~\eqref{eq:GK-mid} are equivalent to $\tfk\rightarrow 0$ and $\tsys \rightarrow\infty$ respectively.
%It can be seen that the approximated GK integral $\Lambda^\text{mid}(t)$ has only one timescale $\tslip$ and has a magnitude of $\lambda_0$.
%In fact, $\Lambda(t)$ is almost equal to $\Lambda^\text{mid}(t)$ in the range of $\tfk \ll t \ll \tsys$, see Fig.~\ref{fig:GK_model}.
%\begin{figure}
%    \centering
%    \includegraphics[width=0.5\linewidth]{figs/fig1_Lambda.pdf}
%    \caption{Comparison of GK integrals $\Lambda(t)$, blue: theoretical solution Eq.~\eqref{eq:GK-mid} and orange: approximate solution Eq.~\eqref{eq:GK-full}.}
%    \label{fig:GK_model}
%\end{figure}
%The blue line is GK integral described in Eq.~\eqref{eq:GK-full} with Maxwell type friction kernel of Eq.~\eqref{eq:friction-Maxwell}.
%The orange line is approximated solution of Eq.~\eqref{eq:GK-mid}.
%\red{
Note that the solution Eq.~\eqref{eq:GK-mid} indeed reproduces 
the plateau feature if the slip length $b$ and the system height $h$ are large enough.
%, $\Lambda(t)$ has a plateau value of $\lambda_0(=\Lambda^\text{mid}(0))$ because of $t/\tslip=0$.
%This is consistent with BB method \cite{Bocquet1994} of %Eq.~\eqref{eq:GK_BB}.
%}
%\par
%If $b$ is not large enough, plateau value of Green-Kubo integral $\Lambda(t)$ is difficult to determine \cite{??}.
Taking into account that Eq.~\eqref{eq:GK-mid} should hold under the condition of
%that the three timescales $\tfk$, $\tslip$ and $\tsys$ can be well separated as:
%
\begin{equation}\label{eq:time_condition_fit}
    \tfk  \ll  t  \ll \tsys,
\end{equation}
we propose a calculation method of $\lambda_0$ as the FC in Eq.~\eqref{eq:navier} by fitting the GK integral with Eq.~\eqref{eq:GK-mid} in the time range $t$ around $\tslip$ timescale satisfying Inequality~\eqref{eq:time_condition_fit}.
Indeed, this time separation is supposed to be the key of the understanding of the S-L friction.\cite{Nakano2019,Nakano2019a} 
%
%\red{
Compared to the method by Hansen et al.,\cite{Hansen2011,Varghese2021} in which
the friction kernel $\lambda(t)$ was assumed to be a Maxwell-type one, % in Eq.~\eqref{eq:friction-Maxwell}, 
the present method does not limit the function form of $\lambda(t)$. %by assuming the timescale separation in  Eq.~\eqref{eq:time_condition_fit}.
In addition, \citet{Nakano2020} proposed a similar expression based on the time separation as in the present study;
however, we used a different framework starting from the Stokes equation for the fluid and the Langevin 
equation for a wall as shown in Appendix~\ref{appsubsec:Lang-Stokes},\cite{Oga2021} and the 
resulting expression for the GK integral is different.
%}
% \red{not needed?: v
% Note that the fitting range cannot be determined uniquely, because $\tslip$ depends on the FC of fitting parameters.
% }
%

%
%
\section{Simulation}
\label{sec:simulation}
All the simulations were performed using LAMMPS.\cite{THOMPSON2022108171} %\cite{Plimpton1995} ( \url{https://doi.org/10.1016/j.cpc.2021.108171}).
We considered a general Lennard-Jones (LJ) liquid confined between parallel walls, see Fig.~\ref{fig:GKinteg_fit}(a),
%
% We used two types of walls, flat and grooved surfaces.
% We used fcc walls exposing a (001) face to the liquid; first neighbors in the solids were bound by a harmonic potential $\Phi_\text{h} (r) = k/2\,(r-r_\text{eq})^2$, with $r$ the interparticle distance, $r_\text{eq} = 0.277$\,nm, and $k = 46.8$\,N/m. 
% We used 8 layers for the flat surface wall and added grooves of $4\times 4$ layers to the wall at regular intervals for the grooved surface wall.
% The grooved wall is uniform in the $y$-direction.
where we used fcc crystal walls composed of 8 atomic layers exposing a (001) face to the liquid;
the first neighbors in the solid particles denoted by s were bound by a harmonic potential:
\begin{equation}
    \Phi_\text{h}^\mathrm{ss} (r_{ij}) 
    = 
    \frac{k^\mathrm{ss}}{2}
    (r^\mathrm{ss}_{ij}-r^\mathrm{ss}_\text{eq})^2,
\end{equation}
with $r^\mathrm{ss}_{ij}$ the interparticle distance between neighboring solid particles $i$ and $j$, $r^\mathrm{ss}_\text{eq} = 0.277$\,nm, and $k^\mathrm{ss} = 46.8$\,N/m.
Interactions between fluid particles (ff) and between fluid and solid particles (fs) were modeled by a 12-6 LJ pair potential:
\begin{equation}\label{eq:LJ}
    \Phi_\text{LJ}^{\alpha\beta} (r^{\alpha\beta}_{ij}) 
    = 
    4\varepsilon^{\alpha\beta}
    \left[
    \left(\frac{\sigma^{\alpha\beta}}{r_{ij}^{\alpha\beta}}\right)^{12} - \left(\frac{\sigma^{\alpha\beta}}{r_{ij}^{\alpha\beta}}\right)^6
    + c_2^{\alpha\beta} \left(\frac{r_{ij}^{\alpha\beta}}{r_\mathrm{c}^{\alpha\beta}}\right)^2 + c_0^{\alpha\beta}
    \right],
\end{equation}
where $r^{\alpha\beta}_{ij}$ is the distance between particle $i$ and $j$, with $\alpha \beta$ being ff or fs. 
This LJ interaction was truncated at a cutoff distance of $r_\mathrm{c}^\mathrm{\alpha\beta} = 3.5\sigma^\mathrm{ff}$,
where the potential 
%\red{
$\Phi_\text{LJ}^{\alpha\beta} (r^{\alpha\beta}_{ij})$
%} 
and the interaction force
%\red{
$-\frac{\mathrm{d}\Phi_\text{LJ}^{\alpha\beta} (r^{\alpha\beta}_{ij})}{\mathrm{d} r^{\alpha\beta}_{ij}}$
%} 
smoothly vanished at $r_\mathrm{c}^{\alpha\beta}$ by adding a quadratic function described by constants coefficients $c_{2}^{\alpha\beta}$ and $c_{0}^{\alpha\beta}$.\cite{Nishida2014}  
We used $\sigma^\mathrm{ff} = 0.34$\,nm, $\epsff = 121$\,K$\cdot \kB$, $\sigma^\mathrm{fs} = 0.345$\,nm, and $\epsfs$ was varied between $0.155 \epsff$ and $0.464 \epsff$ to change the wettability.
The contact angle is $136^\circ$ for $\epsfs=0.155 \epsff$,
$79^\circ$ for $\epsfs=0.310 \epsff$,
and complete wetting for $\epsfs = 0.464 \epsff$.\cite{Oga2019,Ogawa2019} 
The atomic masses of fluid and wall particles were $m_\mathrm{f} = 39.95$\,u and $m_\mathrm{s} = 195.1$\,u.
We used periodic boundary conditions along the surface lateral $x$ and $y$  directions with a box size $L_x = L_y = 6.27$\,nm. 
%\red{, which was sufficiently large to avoid unfavorable box size effects on the friction coefficient.\cite{Ogawa2019}} 
Numbers of fluid and wall particles were 6400 and 8192 respectively,
and the total system height including walls along the surface normal $z$ direction was about 12\,nm.
The distance between the walls was determined by a pressure controlled pre-calculation of 20\,ns in which
an external force equivalent to target pressure 4\,MPa was applied to the outermost layer of the top wall.

We compared the GK measurements by EMD and a reference NEMD (Couette) measurement of the friction coefficient.
For the NEMD system, the outermost layer of the top and bottom walls have constant velocities $\uw^\mathrm{top}=\uw$ and $\uw^\mathrm{bot}=-\uw$ with $u_\mathrm{w}=10$\,m/s.
%\red{
Note that the present shear rate with this setting is in the linear response regime.\cite{Kannam2011,Omori2019} 
%} 
The temperature of the system was set to 100\,K by applying a Langevin thermostat to the 2nd outermost layer of walls
in the $xyz$-direction for the EMD and in the $yz$-direction excluding the shear direction for the NEMD system.
We integrated the equation of motion using the velocity-Verlet algorithm, with a time step of 5\,fs.
The simulation time was 200\,ns. 

\section{Results and discussion
\label{sec:resdis}}
\par
\begin{table}[t]
    \caption{Methods to evaluate the friction coefficient $\lambda_{0}$ in Eq.~\eqref{eq:navier} tested in this study.}
    \centering
    \begin{tabular}{c|c|c|c|c}
        \hline\hline
        symbol 
        & 
        \begin{tabular}{c}
        %\red{
        expression for $\lambda_{0}$
        %}
        \\
        %\red{
        or for $\Lambda(t)$, fitting func.
        %}
        \end{tabular}
        & 
        timescale
        \footnote{$\tfk$, $\tslip$ and $\tsys$ are defined in Eqs.~\eqref{eq:friction-Maxwell}, \eqref{eq:def_tslip} and \eqref{eq:def_tsys}, respectively.}
        & 
        fit. parameters 
        &
        system 
\\ \hline\hline
        $\lambda_\text{mid}$ (present)
        &
        %$\begin{matrix}
        %\lammid \approx \lambda_{0}\mathrm{\ with}
        %\\
        %\lambda_0 \exp\braced{\frac{t}{\tslip}} \text{erfc} \braced{\sqrt{\frac{t}{\tslip}}}
        %\footnote{$\lammid(t)$ in Eq.~\eqref{eq:GK-mid}} 
        %\end{matrix}$
        %\red{
        $\begin{matrix}\text{fit. func.: } \Lambda_\text{mid}(t) = \\
    \lambda_0 \exp\braced{\frac{t}{\tslip}} \text{erfc} \braced{\sqrt{\frac{t}{\tslip}}}  \footnote{Eq.~\eqref{eq:GK-mid}.}\end{matrix}$
    %}
        &
        $t$ 
        %\red{
        $\sim$
        %} 
        $\tslip$ 
        &
        $\lambda_0$ and $\tslip$ 
        & 
        EMD 
\\ \hline
        $\lamshort$ 
        &
        %$\begin{matrix}
        %\lamshort \approx \lambda_{0}\mathrm{\ with}
        %\\
        %\lambda_0(e^{-t/t_1} - e^{-t/t_2})
        %\footnote{$\lamshort(t)$ in Eq.~\eqref{eq:GK-micro} \cite{Oga2019}}
        %\end{matrix}$
        %\red{
        $\begin{matrix}\text{fit. func.: } \Lamshort(t) = \\
    \lambda_0(e^{-t/t_1} - e^{-t/t_2})  \footnote{Eq.~\eqref{eq:GK-micro}.\cite{Oga2019}}\end{matrix}$
    %} 
        & 
        $t$ 
        %\red{
        $\sim$
        %} 
        $\tfk$
        &
        $\lambda_{0}$, $t_{1}$ and $t_{2}$
        &  
        EMD 
 \\ \hline
        $\lammax$ &
        $\lambda_{0} \approx \max\left[\Lambda(t)\right]$ 
        \footnote{Eq.~\eqref{eq:lambda0=Lambdamax}.}
        &
        $t$ 
        %\red{
        $\sim$
        %} 
        $\tfk$ &
        - &
        EMD 
\\ \hline
        $\lamtail$
        & 
        $\displaystyle 
        \lambda_{0} \approx 
        \frac{2\Lambda(\infty)}{1-\frac{h}{\eta}\Lambda(\infty)}$
        \footnote{Eq.~\eqref{eq:Lambda_conv}. 
        %\red{
        Viscosity $\eta$ and liquid hydrodynamic height $h$ must be additionally calculated.
        %}
        }
        & 
        $t>\tsys$
        &
        -
        & EMD 
\\ \hline \hline
        $\lamnemd$ 
        & 
        - 
        & 
        -
        &
        -
        & 
        NEMD
        \footnote{Shearing the walls with $\uw^\mathrm{top} - \uw^\mathrm{bot}=20$\,m/s.}
        \\ \hline\hline
    \end{tabular}
    \label{tab:fric_methods}
\end{table}
To test the present method, we compared five methods to evaluate  $\lambda_{0}$, 
% \begin{enumerate}
%     \item
%     $\lammid$:\ the proposed method in the present study by fitting $\Lambda(t)$ by Eq.~\eqref{eq:GK-mid}.
%     \item
% \end{enumerate}
listed in Table~\ref{tab:fric_methods}:  $\lammid$ as the present one, and four other methods $\lammax$, $\lamshort$, $\lamtail$, and $\lamnemd$. 
The former four are obtained from the GK integral $\Lambda(t)$ in EMD, where 
$\lammid$ and $\lamshort$ are calculated through 
the fitting of functions to $\Lambda(t)$ for the 
corresponding timescale ranges, whilst $\lamnemd$ is evaluated from NEMD simulation of steady-state Couette-type flows through the direct measurement of $\tauw$ and $u_\mathrm{slip}$ in Eq.~\eqref{eq:navier}.\cite{Nakaoka2015,Nakaoka2017,Nakaoka2017b,Oga2019,Ogawa2019} 
%
%
% Each calculation method is explained in Table~\ref{tab:fric_methods}.
%
\par
\begin{figure}
    \centering
    \includegraphics[width=\linewidth]{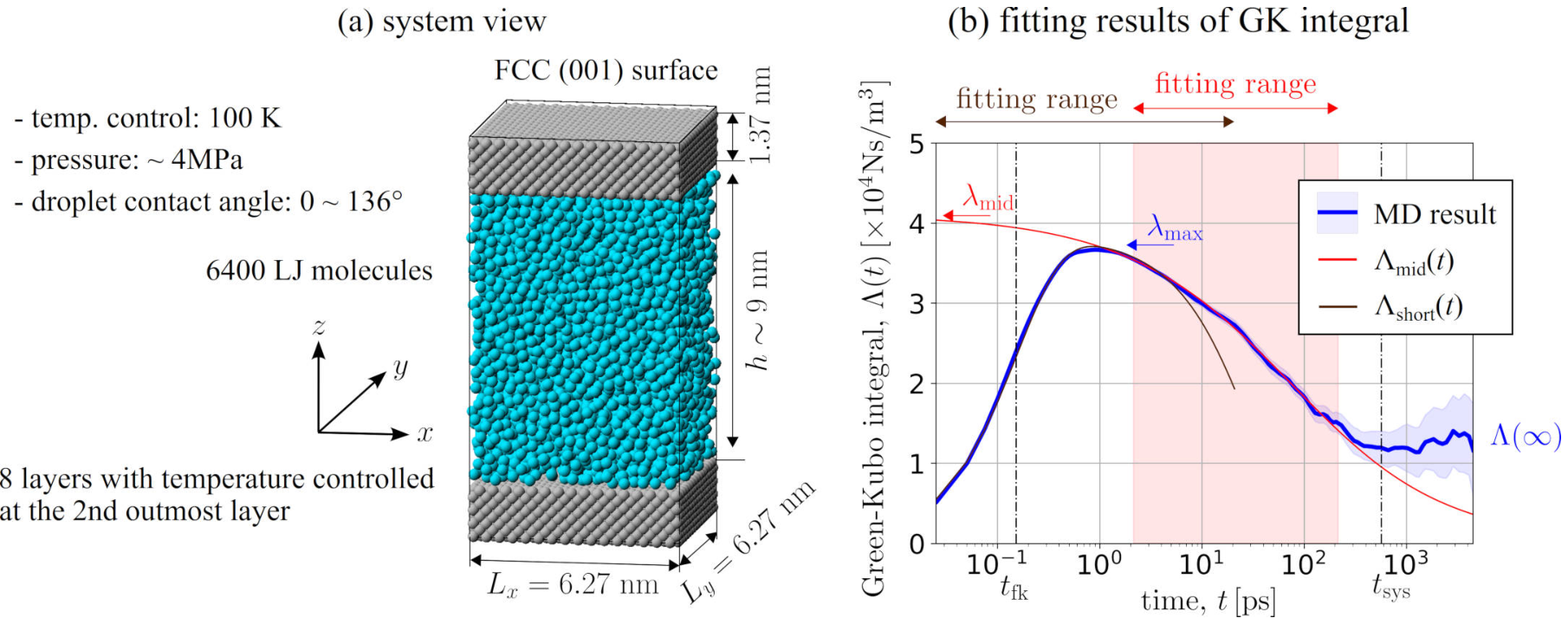}
    \caption{(a) MD simulation system of LJ liquid between flat fcc walls.
    (b) (Blue) GK integral $\Lambda(t)$ for 
    $\epsfs = 0.310 \epsff$, and 
    fitting curves for (red) $\Lammid(t)$ with Eq.~\eqref{eq:GK-mid}, 
    and 
    (brown) $\Lamshort(t)$ with Eq.~\eqref{eq:GK-micro}. 
    %\red{
    The maximum value $\lammax$ as well as $\Lambda(\infty)$ used to compute $\lamtail$ are also indicated. Finally, the timescales $t_\mathrm{fk}$ and  $t_\mathrm{sys}$ are represented on the time axis.
    %} 
%    \red{(c) not needed: Table I is enough?)}%(c) Symbols and simple explanations of obtained FC.
    }
    \label{fig:GKinteg_fit}
\end{figure}
Figure~\ref{fig:GKinteg_fit}~(b) illustrates the four EMD calculation methods 
of the FC from the GK integrals $\Lambda(t)$ listed in Table~\ref{tab:fric_methods} for a system with $\epsfs = 0.310 \epsff$. %as shown in Fig.~\ref{fig:GKinteg_fit} (c).
%
%Fig.~\ref{fig:GKinteg_fit} (b) shows the results of the GK integral $\Lambda(t)$ and FCs when $\epsfs=0.310 \epsff$.
%
For $\lambda_\text{mid}$, we fitted the GK integral with Eq.~\eqref{eq:GK-mid} as the red curved-line in Fig.~\ref{fig:GKinteg_fit}~(b).
The set of fitting parameters is $(\lambda_0,\,\tslip)$,
and the fitting range is $(t^\text{fit}_\text{mid},\,10t^\text{fit}_\text{mid})$, 
where 10 values for $t^\text{fit}_\text{mid}$ were tested in the range of 2.14\,ps and 21.4\,ps at equal log-scale intervals.
For $\lamshort$, we fitted the GK integral $\Lambda(t)$ by
\begin{equation}\label{eq:GK-micro}
\Lamshort(t) = \lambda_0\left(
e^{-\frac{t}{t_1}}-e^{-\frac{t}{t_2}}
\right) %,\quad \red{\lamshort\equiv \lambda_{0}}
\end{equation}
as proposed in our previous study \cite{Oga2019} as the brown line 
 in Fig.~\ref{fig:GKinteg_fit}~(b), 
where the set of fitting parameters is $(\lambda_0,\,t_1,\,t_2)$
and the fitting range is $(0,\, t^\text{fit}_\text{short})$  
with  10 different values of $t^\text{fit}_\text{short}$ tested in the range of 2.14\,ps and 21.4\,ps at equal log-scale intervals.
As observed in Fig.~\ref{fig:GKinteg_fit}~(b), the red and brown fitting lines corresponding to $\Lammid(t)$ in Eq.~\eqref{eq:GK-mid} and $\Lamshort(t)$ in Eq.~\eqref{eq:GK-micro} reproduce $\Lambda(t)$ well for their fitting ranges,  and from these fitting curves, we extract $\lammid$ (displayed with the red arrow) and $\lamshort$. 
In addition, we calculated the other two FC approximations $\lammax$ and $\lamtail$ from the maximum of $\Lambda(t)$ in Eq.~\eqref{eq:lambda0=Lambdamax} and the convergence value $\Lambda(\infty)$ in Eq.~\eqref{eq:Lambda_conv}. 
For simple cases including this example with $\Lambda(t)$ showing a simple behavior of increasing within a short time and decaying after taking the maximum, the former three give similar results, with slight difference with $\lammid, \lamshort > \lambda_\mathrm{max}$; however, we will see later that it is not the case with structured walls.
On the other hand, as mentioned in Sec.~\ref{sec:intro}, 
%\red{
the final plateau value of $\Lambda(t)$
%} 
showed 
large fluctuation as seen in the right-end of the blue curved line Fig.~\eqref{fig:GKinteg_fit}, 
from which $\lamtail$ is evaluated including additional calculation of $\eta$ and $h$ 
obtained in different systems.
\par
%\red{
To determine the liquid height $h$ used to compute $\lamtail$ and $\lamnemd$, 
%On the other hand, 
%for $\lamtail$ and $\lamnemd$, 
we considered that the S-L interface position $z_\mathrm{SL}$ was approximately $\sigfs$ outward the wall surface for the present case,\cite{Herrero2019, Omori2019} and we defined $h$ as: %as mentioned in Sec.~\ref{sec:intro}, we used this position to determine the liquid height $h$ 
% as
\begin{equation}\label{eq:def_h}
    h = z_\mathrm{SL}^\mathrm{top} - z_\mathrm{SL}^\mathrm{bot}. 
\end{equation}
%
%for Eq.~\eqref{eq:GK-full-conv2} to evaluate $\lamtail$, and for 
The slip velocity $\uslip$ used to evaluate  $\lambda_\mathrm{NEMD}$ was then given by: 
\begin{equation}\label{eq:def_uslip}
        \uslip 
        \equiv 
        \frac{1}{2}\left[
        (\uw^\mathrm{top} - \uw^\mathrm{bot})
        -
        h\left(\frac{\ptl u}{\ptl z}\right)_\mathrm{bulk}
        \right]
        = \uw - \frac{h}{2}\left(\frac{\ptl u}{\ptl z}\right)_\mathrm{bulk}
\end{equation}
where $\left(\frac{\ptl u}{\ptl z}\right)_\mathrm{bulk}$ was obtained by 
fitting the average velocity distribution of the liquid bulk with a linear function of $z$.
%}

\begin{figure}
    \centering
    \includegraphics[width=0.5\linewidth]{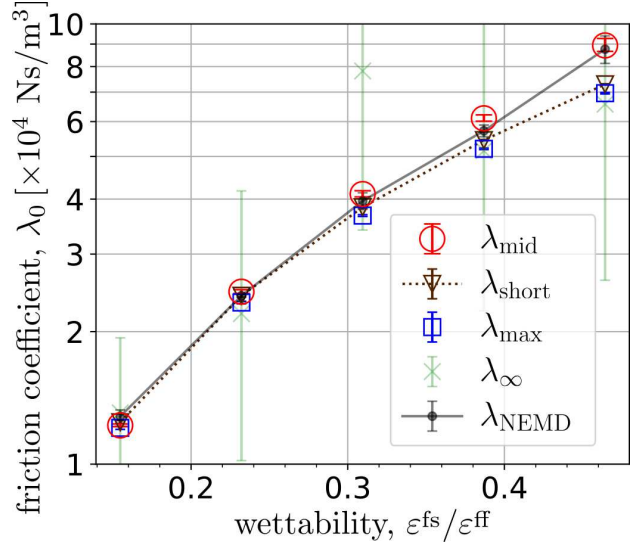}
    \caption{FC for various wettability $\epsfs$.
    The calculation methods for each of the 5 estimates of the FC are shown in Table.~\ref{tab:fric_methods}.
%    \red{$\lambda_{tail}\rightarrow\lamtail$}
    }
    \label{fig:friccoef_homo}
\end{figure}
Figure~\ref{fig:friccoef_homo} shows the FCs obtained for various wettabilities --controlled through $\epsfs$, 
%The FC $\lambda_\text{conv}$ is obtained from the convergence value of the GK integral by using Eq.~\eqref{eq:Lambda_conv}.
where we calculated $\lammid$ and $\lamshort$ as the average for various fitting ranges. The error bars show the uncertainties due to the fitting range and due to the fluctuation of the 
%\red{
GK integral
%} 
$\Lambda(t)$, 
%\red{
which gives larger uncertainty as $t$ increases
%} \red{
(the former was much smaller than the latter).
%}
%and their error bars were evalulated from the standard deviation.
%We also measured the FC $\lambda_\text{NEMD}$ in the corresponding NEMD (Couette) system.
%Here, $\lambda_\text{NEMD}$ was calculated from Navier B.C. of Eq.~\eqref{eq:navier}.
%In the Couette flow, the slip velocity $u_\text{slip}$ depends on the S-L interface position which is arbitrary.
%
Regarding the comparison with the reference NEMD value $\lamnemd$, $\lammid$ --proposed in this study-- reproduced $\lamnemd$ the best in comparison with $\lammax$ and $\lamshort$, which overall underestimated $\lambda_0$, especially for larger FC on more wetting surfaces. One can also note that the error bars for $\lamtail$ 
were too large to precisely evaluate $\lambda_{0}$, at least with the present calculation cost.
\par
\begin{figure}
    \centering
    \includegraphics[width=0.5\linewidth]{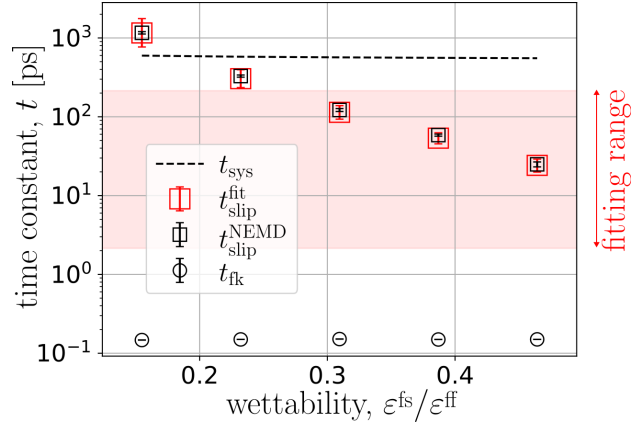}
    \caption{Time scales as a function of wettability;
    $\tsys=\rho h^2/\eta$,
    $\tslip^\text{fit}$  obtained as a fitting parameter of Eq.~\eqref{eq:GK-mid},
    $\tslip^\text{NEMD}=\rho\eta/\lamnemd^2$,
    and $t_\lambda$  obtained by fitting $\langle F_\text{w}(t) F_\text{w}(0) \rangle/S\kB T$
    with the Maxwell-type friction in Eq.~\eqref{eq:friction-Maxwell}.
    }
    \label{fig:time_const}
\end{figure}
%
%\red{---To Appendix??---}
As mentioned in Sec.~\ref{sec:theory},
there are three key timescales $\tfk$, $\tsys$, and $\tslip$ for the GK integral $\Lambda(t)$, and for the present fitting by Eq.~\eqref{eq:GK-mid}, the timescale $\tslip$ 
must be basically separated from $\tfk$ and $\tsys$.
%, \ie $\tslip$ for the present fitting should be between $\tfk$ and $\tsys$.
% The last one, , is the unique time constant of the fitting function $\Lammid(t)$.
To check this separation, we estimated the three timescales in the present systems. 
For $\tfk$, considering that the friction kernel $\lambda(t)$ can be well approximated
by the correlation function in the RHS of Eq.~\eqref{eq:fric_kernel_approx_C_F}
on a short timescale as \cite{Oga2021}
\begin{equation}\label{eq:fric_kernel_approx_C_F}
    \lambda(t) \approx \frac{1}{S\kB T} \langle \Fw(t)\Fw(0)\rangle,
\end{equation}
we fitted the RHS of Eq.~\eqref{eq:fric_kernel_approx_C_F} by the Maxwell-type friction kernel in Eq.~\eqref{eq:friction-Maxwell}. 
We estimated $\tslip$ following two approaches, which we denote
by $\tslip^{\mathrm{mid}}$ and $\tslip^\mathrm{NEMD}$, respectively: 
1) evaluating it as a fitting parameter for $\lammid$ with Eq.~\eqref{eq:GK-mid};  
%as in Fig.~\ref{fig:GKinteg_fit}, 
and 2) calculating $\tslip$ from the FC $\lamnemd$ obtained by NEMD as
\begin{equation}
\label{eq:tslip_NEMD_def}
    \tslip^\mathrm{NEMD} \equiv \frac{\rho \eta}{\lamnemd^{2}}.
\end{equation}
The density and viscosity values $\rho$ and $\eta$ in Eq.~\eqref{eq:tslip_NEMD_def} 
were obtained in the NEMD systems, where $\eta$ was evaluated by %Eq.~\eqref{eq:newtonvisc} as
\begin{equation}
\label{eq:visc_NEMD}
    \eta \equiv \frac{\tauw^\mathrm{NEMD}}{\left(\frac{\ptl u}{\ptl z}\right)_\mathrm{bulk}} , 
\end{equation}
using the average solid-liquid shear force per area $\tauw^\mathrm{NEMD}$ 
measured on the solid surface. These values were also used for the evaluation of 
$\tsys$ given by Eq.~\eqref{eq:def_tsys}. Since the present systems were under pressure 
and temperature control, the resulting $\rho$ and $\eta$ were constant. 
%\red{
On the other hand, $h$ slightly depended on the wettability $\epsfs$ under this condition with a constant number of fluid particles, hence, $t_\mathrm{sys}$ slightly depended on the wettability, too.
%}% %
%As described in Sec.~\eqref{sec:theory}, the separation the three timescales 
%$\tfk$, $\tslip$ and $\tsys$ are the key of the BB integral $\Lambda(t)$ and 
%the present method.
%%\add{
%We estimated $\tslip$ from $\lammid$ and $\lamnemd$ values by 
%%
%\begin{equation}
%    \tslip^\mathrm{mid} \equiv \frac{\rho \eta}{\lammid^{2}},
%    \quad 
%    \tslip^\mathrm{NEMD} \equiv \frac{\rho \eta}{\lamnemd^{2}}.
%\end{equation}
%%
%
%
% The other fitting parameter $\tslip$ was calculated as $\tslip^\text{fit}$
% and compared with the time constant 
% %
% \begin{equation}
% \tslip^\text{NEMD}=\frac{\rho\eta}{\lamnemd^2}
% \end{equation} 
% %
% obtained from FC of NEMD.
%
\par
Figure~\ref{fig:time_const} shows the comparison among the three timescales. 
For the present system, $\tslip$ depended largely on the wettability 
$\epsfs$ as easily imagined from Eq.~\eqref{eq:def_tslip} 
with the results of $\lammid$ in Fig.~\ref{fig:friccoef_homo}. This was 
in contrast to $\tfk$, which was overall below one picosecond and was almost 
independent of the wettability. %$\epsfs$. 
In addition, the 
system timescale $\tsys$ was about several hundreds of picoseconds.
Hence, the timescale separation in Ineq.~\eqref{eq:time_condition_fit}
was well satisfied for $\epsfs \geq 0.3\epsff$; however, 
even for the case of $\epsfs = 0.1\epsff$ where $\tslip$ was  
larger than $\tsys$, the present result $\lammid$ still gave a good estimate 
of $\lamnemd$. 
%\red{
This is probably because of the features of the hyperbolic 
functions $\sinh$ and $\cosh$ in Eq.~\eqref{eq:GK-full}, 
which quickly approach to the 
limit for $h\rightarrow \infty$ in Eq.~\eqref{eq:GK-full-twolimits}.
%}
%\red{---To Appendix??---}
%
%It can be seen that the fitting range shown in orange in Fig.~\ref{fig:time_const} mostly satisfies the conditions of Eq.~\eqref{eq:time_condition_fit}.
%
%As shown in Eq.~\eqref{eq:GK-mid}, the GK integral has a unique time constant $\tslip$
%under the time separation condition of Eq.~\eqref{eq:time_condition_fit},
%but my fitting results reproduce $\tslip$ well even for the case of $\tslip > \tsys$,
%here $\tsys$ is upper limit of the time separation condition.
%}
%
\par
\begin{figure}
    \centering
    \includegraphics[width=\linewidth]{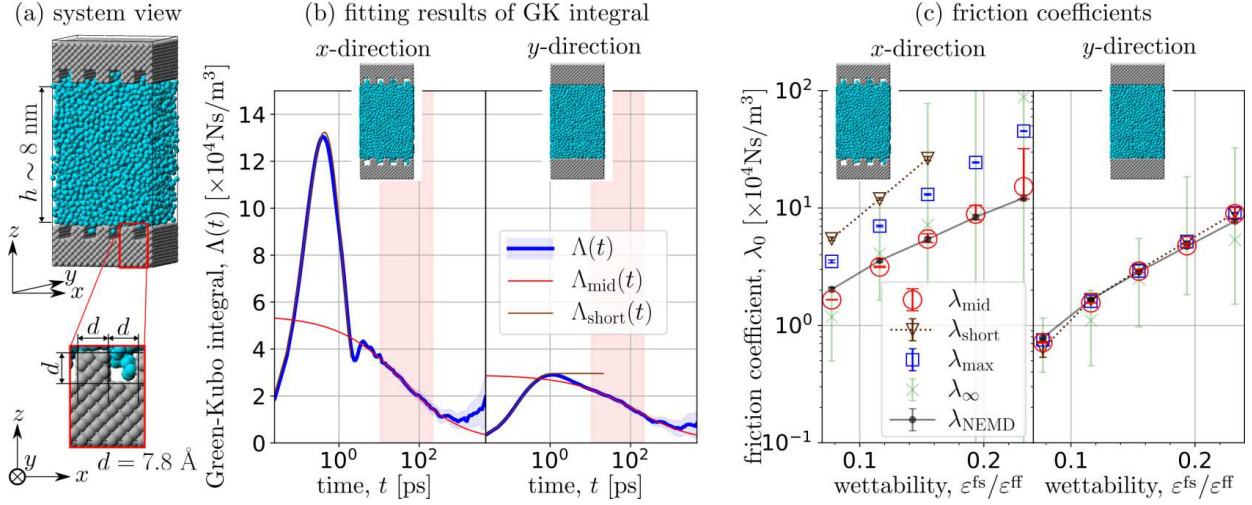}
    \caption{(a) System view with grooved walls 
     for $\epsfs=0.155\epsff$.
    (b) GK integrals $\Lambda(t)$ in the (left) $x$- and (right) $y$-directions 
    and their fitting curves;
    Blue: GK integral, red: fitting curve for $\lammid$, and brown: fitting curve for $\lamshort$. 
%    and 
%    Horizontal lines are FC
%    for different methods.
    (c) Obtained FCs for various wettability $\epsfs$ 
    in the (left) $x$- and (right) $y$-directions; the calculation 
    methods are summarized in Table~\ref{tab:fric_methods}.
%    \red{$\lambda_{tail}\rightarrow\lamtail$}
    }
    \label{fig:GKinteg_fitsys-hetero}
\end{figure}
In addition to the analysis on flat surfaces shown above, 
we applied the FC measurement methods to the S-L friction on 
grooved surfaces as exemplified in Fig.~\ref{fig:GKinteg_fitsys-hetero}(a).
%
%where the groove was parallel to the $y$-direction.
%
For each $\epsfs$, two NEMD simulations with shearing in the $x$- and $y$-directions 
were carried out to obtain $\lamnemd$ considering the heterogeneous 
structure of the surface. 
%\red{
In the following, we will denote $u$ and $v$ the velocity along the $x$- and $y$-directions, respectively.
%}  
%, where the velocity $v$ in the $y$-direction was used instead of $u$. 
Since the density and momentum are inhomogeneous in the $xy$-plane near the 
wall surface, there is no clear definition of the slip velocity or the interface position.
%it is difficult to give a clear definition of the slip velocity or the interface position. Hence, we tentatively defined the slip velocity 
Here we defined the slip velocity
$\uslip$ or $v_\mathrm{slip}$ for the NEMD based on Eqs.~\eqref{eq:def_h} and 
\eqref{eq:def_uslip} as follows.
First, the distributions of mass and momentum density in the $x$- or $y$-direction 
were calculated as the average over the $xy$-plane. 
Second, the velocity distribution 
obtained as momentum density per mass density was fitted only in the bulk part 
with a linear function to obtain $\left(\frac{\ptl u}{\ptl z}\right)_\mathrm{bulk}$ in 
Eq.~\eqref{eq:def_uslip} or $\left(\frac{\ptl v}{\ptl z}\right)_\mathrm{bulk}$
as in the flat wall systems, 
and finally the slip velocity was obtained by extrapolating the fitted velocity 
distribution to the positions 
$z_\mathrm{SL}^\mathrm{top}$ and $z_\mathrm{SL}^\mathrm{bot}$ in Eq.~\eqref{eq:def_h} at 
$\sigma_\text{fs}$ from the top of the grooved wall surface toward the liquid bulk. %, \ie not from the bottom surface of the groove.
%
%On the other hand, 
An EMD simulation was also run for each $\epsfs$ to obtain $\Lambda^{x}(t)$ and 
$\Lambda^{y}(t)$ in the two directions using $\Fw^{x}$ and $\Fw^{y}$ as the S-L 
friction force $\Fw$ in Eq.~\eqref{eq:def_GKint}.
\par
%\red{short comment about the complex curve in $x$-dir physical reason! not hydrodynamic!} 
Figure~\ref{fig:GKinteg_fitsys-hetero}(b) shows an example of the GK integrals 
$\Lambda^{x}(t)$ and $\Lambda^{y}(t)$ for the grooved surface system, with 
$\epsfs = 0.155\epsff$. The GK integrals $\Lambda^{x}(t)$ in the left panel 
have a complex shape, with a sharp peak in the short timescale below a few 
picoseconds, and a slow decay afterwards, which was not observed for 
$\Lambda^{y}(t)$ in the right panel nor $\Lambda(t)$ for the flat wall 
system in Fig.~\ref{fig:GKinteg_fit}(b).
This short-time behavior is due to the local vibration of the fluid 
particles confined in the grooves, and not related to the hydrodynamic motion.
We fitted $\Lamshort(t)$ and $\Lammid(t)$ to $\Lambda^{x}(t)$ 
and $\Lambda^{y}(t)$ shown with brown and red lines to obtain $\lamshort$ and 
$\lammid$ in both directions as well as $\lammax$ summarized in Table~\ref{tab:fric_methods}. 
%
%Even in a heterogeneous wall system, as in the case of flat wall system,
%
%\par
Figure~\ref{fig:GKinteg_fitsys-hetero}(c) shows the comparison between 
the estimated $\lambda_{0}$ in the $x$- and $y$-directions. As imagined 
from the complex GK-integral $\Lambda^{x}(t)$, $\lamshort$ and $\lammax$ 
in the $x$-direction (left panel) using the short timescale resulted 
in much larger estimate than $\lammid$, while the latter corresponded well with the 
%tentative 
NEMD estimate $\lamnemd$. On the other hand, 
%\red{
for the FC in the 
$y$-direction (left panel), all EMD estimates except $\lamtail$ 
reproduced $\lamnemd$ well.
%}
This also indicates that the present 
%\textcolor{blue}
%{
NEMD
%} 
estimate using the 
above-mentioned definition of the slip velocity was reasonable.
Considering the two results, the present FC measurement method $\lammid$ 
properly 
%\textcolor{blue}{
evaluates
%} 
the FC even in this heterogeneous-wall system.
\section{CONCLUDING REMARKS}
In this study, we proposed a method to calculate the solid-liquid FC from EMD simulations
%by using the GK-integral obtained in EMD systems 
of a liquid confined between parallel solid walls, by fitting the GK integral for the timescale range where the GK integral slowly decays with time.
The fitting function was derived from the analytical solution considering that 
the timescales of the friction kernel and bulk viscous dissipation can be separated.
We compared the resulting FCs with those obtained with other EMD-based methods and with NEMD  simulations, for a Lennard-Jones liquid confined between flat crystalline walls as well as between grooved walls with different wettability, and showed that the present method extracts the FC with 
%\red{
excellent
%} 
%sufficient 
accuracy for various systems, with easy implementation and low calculation cost. 

%
%\add{Todo}
%
\begin{acknowledgments}
%We thank Konan Imadate  for fruitful discussion. 
H.O, T.O. and Y.Y. were supported by JSPS KAKENHI grant (Nos. JP21J20580, JP18K03929, 
and JP22H01400), Japan, respectively. 
Y.Y. was also supported by JST CREST grant (No. JPMJCR18I1), Japan.
\end{acknowledgments}
%
%\newline
\vspace{5mm} \par \noindent
\textbf{DATA AVAILABILITY}
%\newline
\par
The data that support the findings of this study are available from the corresponding author upon reasonable request.
\vspace{5mm} \par \noindent
\textbf{AUTHOR DECLARATIONS}
\newline
\textbf{Conflict of Interest}
\par
The authors have no conflicts to disclose.
\appendix
\section{Derivation of a theoretical solution of the GK integral}
\label{appsec:GK-derivation}
\subsection{Derivation through the combination of Langevin equation and Stokes equation}
\label{appsubsec:Lang-Stokes}
We consider a system where a liquid is confined between two solid walls under no external field, where the top wall is fixed.
Let the bottom wall move freely in a wall-tangential direction $x$; its motion can be described by a generalized Langevin equation:\cite{Bocquet2013} 
\begin{equation}\label{eq:langevin}
	M\dd{U}{t}=-S\int_0^t\xi(t-t')U(t') \rd t'+ \Fw(t),
\end{equation}
where $M$, $S$ and $U$ are the mass, the surface area and the $x$-direction velocity of the bottom wall respectively; $\xi$ is the friction kernel and $\Fw$ is the random force that originates from the direct interaction between the solid and liquid particles. Assuming energy equipartition, Eq.~\eqref{eq:langevin} leads to the fluctuation-dissipation theorem:
\begin{equation}\label{eq:rand}
	C_{\Fw}(t) \equiv
	\bracket{%
	\Fw(t)\,
	\Fw(0)
	} = S \kB  T \xi(t).
\end{equation} 
The motion of the liquid in response to the bottom wall motion can be described by the Stokes equation:
\begin{equation}\label{eq:stokes}
	\pd{u(z,t)}{t}=\frac{\eta}{\rho}\pdd{u(z,t)}{z}
\end{equation}
with the Navier boundary condition defined on the top and bottom hydrodynamic boundaries at $z=0$ and $h$ respectively given by
\begin{equation}\label{eq:bc}
	\begin{cases}
		\eta\left.\pd{u(z,t)}{z}\right|_{z=h}=%
		\int_0^t \lambda(t-t')\left[
		-u(h,t')
		\right]\rd t', \\[2mm]
		\eta\left.\pd{u(z,t)}{z}\right|_{z=0}=%
		\int_0^t \lambda(t-t')\left[
		u(0,t')-U(t')
		\right]\rd t' ,
	\end{cases}
\end{equation}
where $u$, $t$, $\rho$, $\eta$ and $\lambda$  denote the liquid velocity in the $x$-direction, the time, the bulk liquid density, the bulk liquid viscosity, and the Navier friction coefficient (FC), respectively. Note that 
%for a wide frequency range \cite{Omori2019}\footnote{We describe later in this SM why we can employ the frequency-independent bulk viscosity in our analysis.}.
the non-Markovian nature is included in $\lambda$.\cite{Omori2019} 
Denoting the Fourier-Laplace transformed variables with tilde as
\begin{equation*}
    \tilde f (\omega) \equiv \int_0^\infty f(t)e^{-i\omega t} \text d t,
    \tag{\ref{eq:def_FL}}
\end{equation*}
%
%Fourier-Laplace transformed, 
the solution of Eqs.~\eqref{eq:stokes} and \eqref{eq:bc} is written in a compact form as $\fl{u}=\fl{\gamma}\fl{U}$ for the liquid velocity on the bottom wall with
\begin{equation}
	\fl{\gamma}
	=%
	\frac{%
	\fl{\lambda}
	\left[
	\fl{\lambda} \sinh(\zeta h) + 
	\eta\zeta \cosh(\zeta h)
	\right]
	}{%
	(\fl{\lambda}^2 + \eta^2 \zeta^2)\sinh(\zeta h) + 
	2 \fl{\lambda}\eta\zeta \cosh(\zeta h)
	}.
\end{equation}
Because the first term on the right hand side of Eq.~\eqref{eq:langevin} can also be rewritten as
\begin{equation}
    -S\int_0^t\xi(t-t')U(t')\rd t'  = -S\int_0^t \lambda(t-t')[U(t')-u(0,t')]\rd t'
\end{equation}
in terms of the slip velocity on the wall, the friction kernel $\xi$ can be written as 
\begin{equation}
\fl{\xi}=\fl{\lambda}(1-\fl{\gamma}).
\end{equation}
Combined with Eq.~\eqref{eq:rand}, the expression for the force autocorrelation function writes 
\begin{equation*}\label{eq_app:GK-full}
	\frac{\fl{C}_{\Fw}}{Sk_B T}
	=%
	\frac{%
	\fl{\lambda} \eta\zeta
	\left[
	\eta\zeta \sinh(\zeta h) + 
	\fl{\lambda} \cosh(\zeta h)
	\right]
	}
	{%
	(\fl{\lambda}^2 + \eta^2 \zeta^2)\sinh(\zeta h) + 
	2 \fl{\lambda}\eta\zeta \cosh(\zeta h)
	},
	\tag{\ref{eq:GK-full}}
\end{equation*}
%Here the tilde indicates that the variables are Fourier-Laplace transformed and 
where $\zeta$ is given by
\begin{equation*}
\label{eq_app:def_zeta} 
\zeta = \sqrt{\frac{i\rho\omega}{\eta}}
\tag{\ref{eq:def_zeta}}
\end{equation*}
as a function of the angular frequency $\omega$. 
\subsection{Asymptotic behavior}
\label{appsubsec:Lambda_infty}
\par
% Equation~\eqref{eq:GK-full} in the FL-form as well as including $\zeta$ in Eq.~\eqref{eq:def_zeta} is rather complex and the solution does not give clear outlook 
% of the physical aspect of the solid-liquid friction. More practically, calculating the FC of 
% $\lambda_{0}$ in the Navier BC in Eq.~\eqref{eq:navier}
% directly from Eq.~\eqref{eq:GK-full} is not trivial.
% Regarding this point, 
As one of possible methods to obtain $\lambda_{0}$, we proposed to use the convergence value $\Lambda(\infty)$ in our previous study,\cite{Oga2021} 
in which we used the following relations:
\begin{equation}\label{eq_app:final_theorem}
    \lim_{\omega\rightarrow 0}
    \frac{%
        \fl{C}_{\Fw}(\omega)
    }
    {%
        S\kB T
    }
    =
    \lim_{\omega\rightarrow 0} \int_{0}^{\infty}
    \frac{%
        C_{\Fw}(t)
    }
    {%
        S\kB T
    }
    e^{-i \omega t} dt 
    =
    \lim_{t \rightarrow \infty}\Lambda(t),
\end{equation}
and 
\begin{equation}\label{eq_app:lambda0_FL}
    \lim_{\omega\rightarrow 0} \tilde{\lambda}(\omega)
    =
    \lim_{\omega\rightarrow 0} \int_{0}^{\infty}\lambda(t)e^{-i \omega t} dt 
    =
    \lambda_{0}.
\end{equation}
%
%
%considering that 
By inserting Eqs.~\eqref{eq_app:GK-full} and \eqref{eq_app:def_zeta} into Eq.~\eqref{eq_app:final_theorem},
it follows for the convergence value of the GK integral that 
\begin{align}\label{eq_app:GK-full-conv}
    \lim_{t\rightarrow\infty} \Lambda(t)
    &= 
    % \lim_{\omega\rightarrow 0}
    % \frac{%
    % \fl{C}_{\Fw}(\omega)
    % }{%
    % S\kB T
    % } 
    \lim_{\omega\rightarrow 0}
    \frac{%
    	\fl{\lambda} \eta\zeta
    	\left[
        	\eta\zeta \sinh(\zeta h) + 
        	\fl{\lambda} \cosh(\zeta h)
    	\right]
    }{%
    	(\fl{\lambda}^2 + \eta^2 \zeta^2)\sinh(\zeta h) + 
    	2 \fl{\lambda}\eta\zeta \cosh(\zeta h)
    }
    \notag \\
    &= \lim_{\omega\rightarrow 0}
    	\frac{%
            	\eta\zeta \tanh(\zeta h) + 
            	\fl{\lambda}
    	}
    	{%
        	\left(\frac{\fl{\lambda}}{\eta \zeta} + \frac{\eta \zeta}{\fl{\lambda}}\right)\tanh(\zeta h) + 
        	2 
    	}.
\end{align}
By considering Eq.~\eqref{eq:def_zeta} and 
\begin{equation}
    \lim_{\zeta h\rightarrow 0} \frac{\tanh (\zeta h)}{\zeta h} = 1,
\end{equation}
Eq.~\eqref{eq_app:GK-full-conv} results in
\begin{align*}\label{eq_app:GK-full-conv2}
\lim_{t\rightarrow\infty} \Lambda(t)
    &= \lim_{\omega\rightarrow 0}
    	\frac{%
            	\eta\zeta \cdot \zeta h \frac{\tanh(\zeta h)}{\zeta h} + 
            	\fl{\lambda}
    	}
    	{%
        	\left(\frac{\fl{\lambda}}{\eta \zeta} + \frac{\eta \zeta}{\fl{\lambda}}\right)
        	\zeta h\frac{\tanh(\zeta h)}{\zeta h} + 
        	2 
    	}
% 	\notag \\
%     &= \lim_{\omega\rightarrow 0}
%     	\frac{%
%             	\eta h \zeta^2 \frac{\tanh(\zeta h)}{\zeta h}  + 
%             	\fl{\lambda}
%     	}{%
%         	\left(\frac{h}{\eta}\fl{\lambda} + \frac{\eta h}{ \fl{\lambda}}\zeta^2 \right)
%           	\frac{\tanh(\zeta h)}{\zeta h} 
%           	+ 2
%     	}
	\notag \\
	&= \frac{\lambda_0}{\frac{h}{\eta}\lambda_0+2}
	= \frac{\lambda_0}{\frac{h}{b}+2},
	\tag{\ref{eq:GK-full-conv2}}
\end{align*}
%
%\red{14--18 to Appendix}
%
where Eq.~\eqref{eq:def_b} is used for the final equality.
Hence, from Eqs.~\eqref{eq:def_b} and \eqref{eq_app:GK-full-conv2} $\lambda_0$ can be evaluated  by
\begin{equation}
\label{eq_app:Lambda_conv}
    \lambda_0 = \frac{2\Lambda(\infty)}{1-\frac{h}{\eta}\Lambda(\infty)}.
\end{equation}
\section{velocity distribution of Couette flow system}
\begin{figure}
    \centering
    \includegraphics[width=\linewidth]{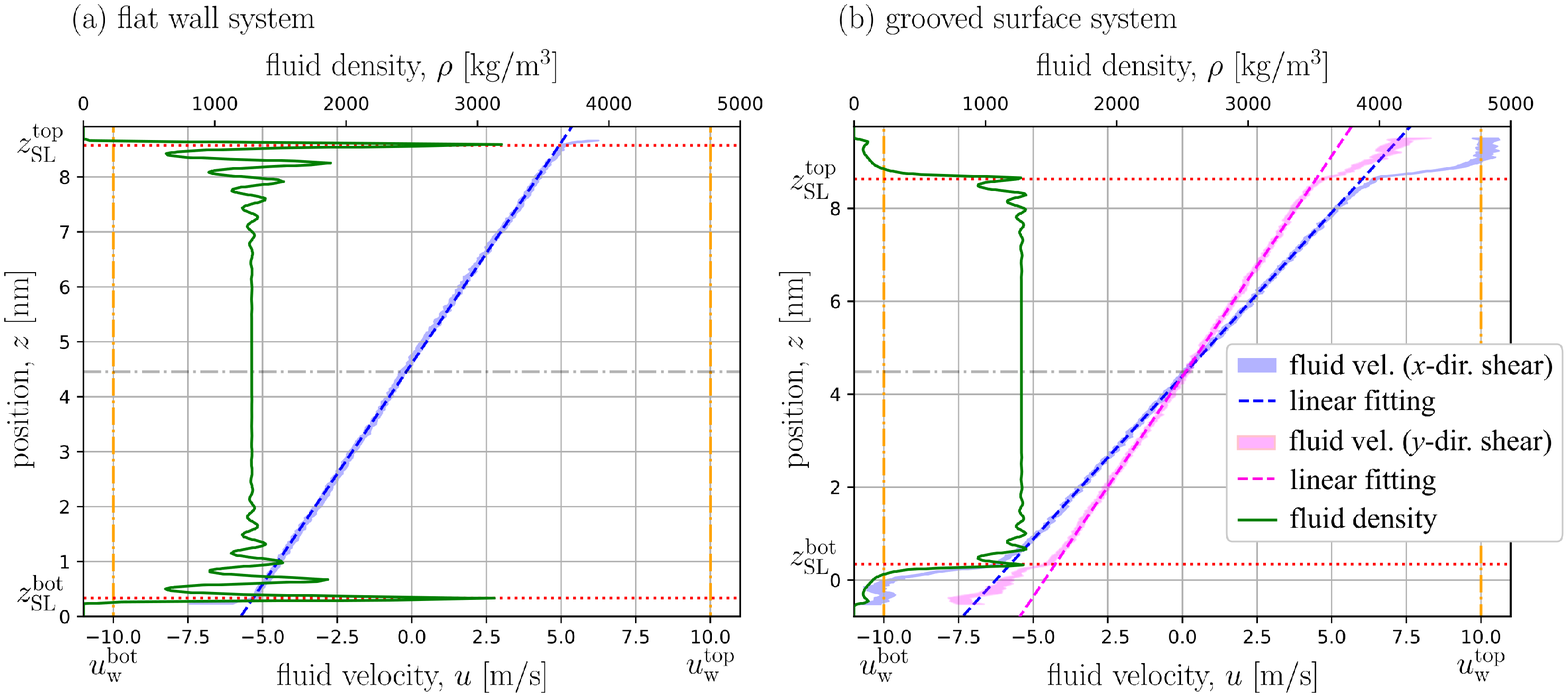}
    \caption{
     Distributions of the fluid velocity and density for 
    (a) flat wall system with $\epsfs=0.310\epsff$, 
    and 
    (b) grooved surface system with $\epsfs=0.155\epsff$.
    Blue: (solid) fluid velocity distribution for $x$-direction shear and (dashed) linear fitting.
    Magenta: (solid) fluid velocity distribution for $y$-direction shear and (dashed) linear fitting.
    Green: fluid density.
    Red dotted lines: positions of the top and bottom solid-liquid interfaces
    $z^{\mathrm{top}}_{\mathrm{SL}}$ and $z^{\mathrm{bot}}_{\mathrm{SL}}$.
    Orange dashed lines: wall velocities 
    $u_\mathrm{w}^\mathrm{bot}$ and $u_\mathrm{w}^\mathrm{top}$.
    }
    \label{fig:velocity_distribution}
\end{figure}
Non-equilibrium MD (NEMD) simulations of Couette-type flow were performed to compute the friction coefficient (FC) $\lamnemd$.
Figure ~\ref{fig:velocity_distribution} shows examples of the distributions of the fluid velocity and density, where two Couette-type MD simulations with shear in the $x$- and $y$-directions were carried out for each wettability parameter $\epsfs$ by moving the top and bottom walls in opposite directions ($\pm x$- or $\pm y$-directions), whilst single simulation ($x$-shear) was run for a flat wall system considering the symmetry of the system.
%Since the surface is not supposed to be isotropic % uniformsymmetrical in the $x$ and $y$ directions
The velocity distributions with shear in the $x$- and $y$-directions for the grooved surface systems are shown in blue and pink, respectively in Fig.~\ref{fig:velocity_distribution}(b). 
Indeed, the velocity and density distributions were not completely quasi-one-dimensional, and the streamline was not parallel to the shear direction for the grooved surface system especially around the wall, \eg the velocities at two points above the concave and convex regions with the same $z$-coordinate were different, but the velocity inhomogeniety in the $xy$-plane quickly vanished and the time-averaged velocity distribution away from the wall was considered to be quasi-one-dimensional. Considering that the present framework is based on the one-dimension Stokes equation as described in Appendix~\ref{appsec:GK-derivation}, we averaged the physical quantities in the $xy$-plane to extract $\lamnemd$ in this study.
The positions of the top and bottom solid-liquid interfaces
$z^{\mathrm{top}}_{\mathrm{SL}}$ and $z^{\mathrm{bot}}_{\mathrm{SL}}$ 
are indicated by the red lines in the Fig.~\ref{fig:velocity_distribution} (see the definition in the main text).
In the case of Fig.~\ref{fig:velocity_distribution}(b), the slip velocity for the shear in the $x$-direction is smaller than that for the shear in the $y$-direction.
\section{friction kernel}
\begin{figure}[t]
    \centering
    \includegraphics[width=\linewidth]{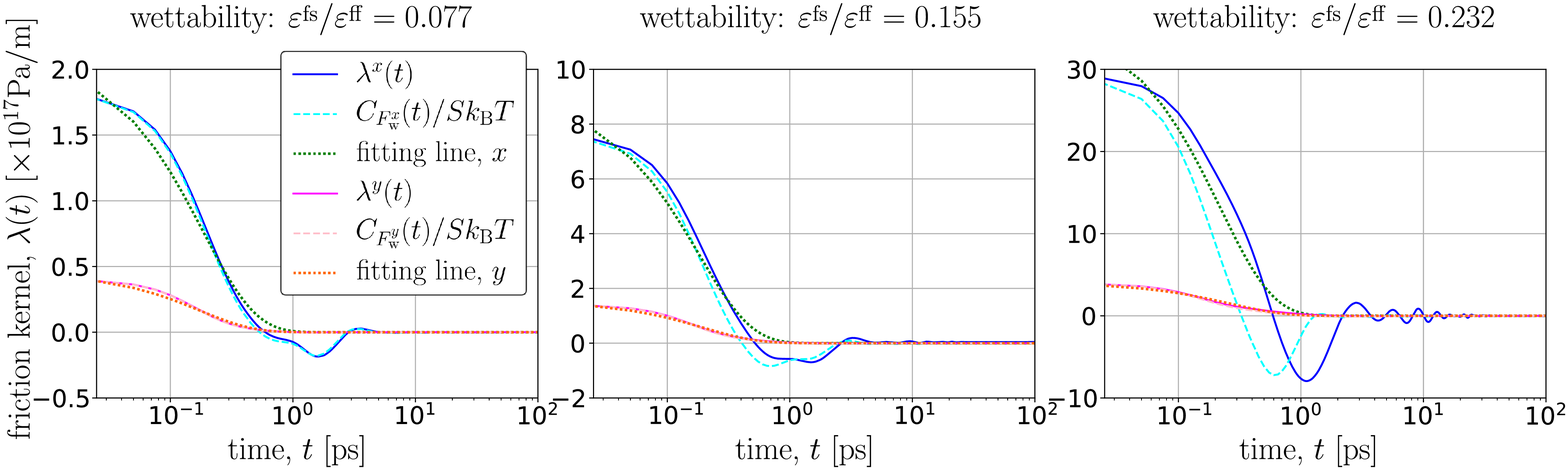}
    \caption{
    Friction kernel $\lambda(t)$ for grooved surface systems with wettability parameter  $\epsfs$ of 
    (left) $0.077\epsff$,
    (center) $0.155\epsff$,
    and 
    (right)  $0.232\epsff$.
    Blue: (solid) friction kernel $\lambda(t)$ and (dashed) $C_{F_{\mathrm{w}}}(t)/S\kB T$ 
    for the $x$-direction.
    Magenta (solid) friction kernel $\lambda(t)$ and (dashed) $C_{F_{\mathrm{w}}}(t)/S\kB T$
    for the $y$-direction.
    Fitting lines with the Maxwell-type kernel in Eq.~\eqref{eq:friction-Maxwell} are shown in green and orange as well.
    }
    \label{fig:friction_kernel}
\end{figure}
To investigate 
%\rem{the cause of} 
the complex shape of GK integral $\Lambda^x(t)$ in Fig.~\ref{fig:GKinteg_fitsys-hetero},
we numerically solved Eq.~\eqref{eq:GK-full} with respect to $\lambda(t)$
for the grooved surface system using the method proposed in our previous study;\cite{Oga2021}
at first, we calculated $\tilde{\lambda}(\omega)$ by numerically solving Eq.~\eqref{eq:GK-full} by using the FL-transform $\tilde{C}_{F_\mathrm{w}}(\omega)$ of the autocorrelation function
${C}_{F_\mathrm{w}}(t)$, 
and then, we obtained $\lambda(t)$ by performing the inverse-FL transform of $\tilde{\lambda}(\omega)$.
Figure~\ref{fig:friction_kernel} shows the results of $\lambda(t)$ 
obtained for (blue) $x$- and (magenta) $y$-directions.
The friction kernel $\lambda(t)$ corresponded well with $\frac{C_{F_\mathrm{w}}(t)}{S\kB T}$ shown in cyan and pink in the $x$- and $y$-directions, respectively for small friction coefficients, \ie  for low wettability.
This is consistent with the fact that Eq.~\eqref{eq:fric_kernel_approx_C_F} holds when
\begin{equation}
\sqrt{\frac{\eta t}{\rho}}\ll \min \left(
\left|\frac{\eta}{\tilde{\lambda}}\right|, h\right)
\end{equation}
is satisfied.\cite{Oga2021} 
We also attempted to fit the kernel by the Maxwell type kernel in Eq.~\eqref{eq:friction-Maxwell} as shown with green and orange lines. The kernel $\lambda(t)$ can be well approximated by the Maxwell type kernel for the $y$-direction, whilst the complex kernel shape cannot be properly expressed by the Maxwell type kernel for the $x$-direction.
The oscillating behavior of the friction kernel $\lambda(t)$ observed  for $\epsfs=0.232 \epsff$ in the $x$-direction within a short timescale $< 10$~ps is supposed to be due to the vibration of the fluid particles confined in the groove.
%\red{
Indeed, this complex kernel shape $\lambda^{x}(t)$ 
%largely affects 
%\blue{
is reflected in
%}
the complex GK integral $\Lambda^{x}(t)$ in Fig.~\ref{fig:GKinteg_fitsys-hetero}.
%}

%
%
%\bibliographystyle{jcp}
%\bibliography{./reference.bib}

\begin{thebibliography}{48}%
\makeatletter
\providecommand \@ifxundefined [1]{%
 \@ifx{#1\undefined}
}%
\providecommand \@ifnum [1]{%
 \ifnum #1\expandafter \@firstoftwo
 \else \expandafter \@secondoftwo
 \fi
}%
\providecommand \@ifx [1]{%
 \ifx #1\expandafter \@firstoftwo
 \else \expandafter \@secondoftwo
 \fi
}%
\providecommand \natexlab [1]{#1}%
\providecommand \enquote  [1]{``#1''}%
\providecommand \bibnamefont  [1]{#1}%
\providecommand \bibfnamefont [1]{#1}%
\providecommand \citenamefont [1]{#1}%
\providecommand \href@noop [0]{\@secondoftwo}%
\providecommand \href [0]{\begingroup \@sanitize@url \@href}%
\providecommand \@href[1]{\@@startlink{#1}\@@href}%
\providecommand \@@href[1]{\endgroup#1\@@endlink}%
\providecommand \@sanitize@url [0]{\catcode `\\12\catcode `\$12\catcode
  `\&12\catcode `\#12\catcode `\^12\catcode `\_12\catcode `\%12\relax}%
\providecommand \@@startlink[1]{}%
\providecommand \@@endlink[0]{}%
\providecommand \url  [0]{\begingroup\@sanitize@url \@url }%
\providecommand \@url [1]{\endgroup\@href {#1}{\urlprefix }}%
\providecommand \urlprefix  [0]{URL }%
\providecommand \Eprint [0]{\href }%
\providecommand \doibase [0]{https://doi.org/}%
\providecommand \selectlanguage [0]{\@gobble}%
\providecommand \bibinfo  [0]{\@secondoftwo}%
\providecommand \bibfield  [0]{\@secondoftwo}%
\providecommand \translation [1]{[#1]}%
\providecommand \BibitemOpen [0]{}%
\providecommand \bibitemStop [0]{}%
\providecommand \bibitemNoStop [0]{.\EOS\space}%
\providecommand \EOS [0]{\spacefactor3000\relax}%
\providecommand \BibitemShut  [1]{\csname bibitem#1\endcsname}%
\let\auto@bib@innerbib\@empty
%</preamble>
\bibitem [{\citenamefont {Eijkel}\ and\ \citenamefont {{van den
  Berg}}(2005)}]{Eijkel2005}%
  \BibitemOpen
  \bibfield  {author} {\bibinfo {author} {\bibfnamefont {J.~C.}\ \bibnamefont
  {Eijkel}}\ and\ \bibinfo {author} {\bibfnamefont {A.}~\bibnamefont {{van den
  Berg}}},\ }\href {https://doi.org/10.1007/s10404-004-0012-9} {\bibfield
  {journal} {\bibinfo  {journal} {Microfluid. Nanofluidics}\ }\textbf {\bibinfo
  {volume} {1}},\ \bibinfo {pages} {249} (\bibinfo {year} {2005})}\BibitemShut
  {NoStop}%
\bibitem [{\citenamefont {Sparreboom}\ \emph {et~al.}(2009)\citenamefont
  {Sparreboom}, \citenamefont {{van den Berg}},\ and\ \citenamefont
  {Eijkel}}]{Sparreboom2009}%
  \BibitemOpen
  \bibfield  {author} {\bibinfo {author} {\bibfnamefont {W.}~\bibnamefont
  {Sparreboom}}, \bibinfo {author} {\bibfnamefont {A.}~\bibnamefont {{van den
  Berg}}},\ and\ \bibinfo {author} {\bibfnamefont {J.~C.}\ \bibnamefont
  {Eijkel}},\ }\href {https://doi.org/10.1038/nnano.2009.332} {\bibfield
  {journal} {\bibinfo  {journal} {Nat. Nanotechnol.}\ }\textbf {\bibinfo
  {volume} {4}},\ \bibinfo {pages} {713} (\bibinfo {year} {2009})}\BibitemShut
  {NoStop}%
\bibitem [{\citenamefont {Bocquet}\ and\ \citenamefont
  {Charlaix}(2010)}]{Bocquet2010}%
  \BibitemOpen
  \bibfield  {author} {\bibinfo {author} {\bibfnamefont {L.}~\bibnamefont
  {Bocquet}}\ and\ \bibinfo {author} {\bibfnamefont {E.}~\bibnamefont
  {Charlaix}},\ }\href {https://doi.org/10.1039/b909366b} {\bibfield  {journal}
  {\bibinfo  {journal} {Chem. Soc. Rev.}\ }\textbf {\bibinfo {volume} {39}},\
  \bibinfo {pages} {1073} (\bibinfo {year} {2010})}\BibitemShut {NoStop}%
\bibitem [{\citenamefont {Schoch}\ \emph {et~al.}(2008)\citenamefont {Schoch},
  \citenamefont {Han},\ and\ \citenamefont {Renaud}}]{Schoch2008}%
  \BibitemOpen
  \bibfield  {author} {\bibinfo {author} {\bibfnamefont {R.~B.}\ \bibnamefont
  {Schoch}}, \bibinfo {author} {\bibfnamefont {J.}~\bibnamefont {Han}},\ and\
  \bibinfo {author} {\bibfnamefont {P.}~\bibnamefont {Renaud}},\ }\href
  {https://doi.org/10.1103/RevModPhys.80.839} {\bibfield  {journal} {\bibinfo
  {journal} {Rev. Mod. Phys.}\ }\textbf {\bibinfo {volume} {80}},\ \bibinfo
  {pages} {839} (\bibinfo {year} {2008})}\BibitemShut {NoStop}%
\bibitem [{\citenamefont {Sparreboom}\ \emph {et~al.}(2010)\citenamefont
  {Sparreboom}, \citenamefont {{van den Berg}},\ and\ \citenamefont
  {Eijkel}}]{Sparreboom2010}%
  \BibitemOpen
  \bibfield  {author} {\bibinfo {author} {\bibfnamefont {W.}~\bibnamefont
  {Sparreboom}}, \bibinfo {author} {\bibfnamefont {A.}~\bibnamefont {{van den
  Berg}}},\ and\ \bibinfo {author} {\bibfnamefont {J.~C.}\ \bibnamefont
  {Eijkel}},\ }\href {https://doi.org/10.1088/1367-2630/12/1/015004} {\bibfield
   {journal} {\bibinfo  {journal} {New J. Phys.}\ }\textbf {\bibinfo {volume}
  {12}},\ \bibinfo {pages} {015004} (\bibinfo {year} {2010})}\BibitemShut
  {NoStop}%
\bibitem [{\citenamefont {Striolo}\ \emph {et~al.}(2016)\citenamefont
  {Striolo}, \citenamefont {Michaelides},\ and\ \citenamefont
  {Joly}}]{Striolo2016}%
  \BibitemOpen
  \bibfield  {author} {\bibinfo {author} {\bibfnamefont {A.}~\bibnamefont
  {Striolo}}, \bibinfo {author} {\bibfnamefont {A.}~\bibnamefont
  {Michaelides}},\ and\ \bibinfo {author} {\bibfnamefont {L.}~\bibnamefont
  {Joly}},\ }\href {https://doi.org/10.1146/annurev-chembioeng-080615-034455}
  {\bibfield  {journal} {\bibinfo  {journal} {Annu. Rev. Chem. Biomol. Eng.}\
  }\textbf {\bibinfo {volume} {7}},\ \bibinfo {pages} {533} (\bibinfo {year}
  {2016})}\BibitemShut {NoStop}%
\bibitem [{\citenamefont {Navier}(1823)}]{Navier1823}%
  \BibitemOpen
  \bibfield  {author} {\bibinfo {author} {\bibfnamefont {C.}~\bibnamefont
  {Navier}},\ }\href@noop {} {\bibfield  {journal} {\bibinfo  {journal}
  {M{\'e}moires de l’Acad{\'e}mie Royale des Sciences de l’Institut de
  France}\ }\textbf {\bibinfo {volume} {6}},\ \bibinfo {pages} {389} (\bibinfo
  {year} {1823})}\BibitemShut {NoStop}%
\bibitem [{\citenamefont {Falk}\ \emph {et~al.}(2010)\citenamefont {Falk},
  \citenamefont {Sedlmeier}, \citenamefont {Joly}, \citenamefont {Netz},\ and\
  \citenamefont {Bocquet}}]{Falk2010}%
  \BibitemOpen
  \bibfield  {author} {\bibinfo {author} {\bibfnamefont {K.}~\bibnamefont
  {Falk}}, \bibinfo {author} {\bibfnamefont {F.}~\bibnamefont {Sedlmeier}},
  \bibinfo {author} {\bibfnamefont {L.}~\bibnamefont {Joly}}, \bibinfo {author}
  {\bibfnamefont {R.~R.}\ \bibnamefont {Netz}},\ and\ \bibinfo {author}
  {\bibfnamefont {L.}~\bibnamefont {Bocquet}},\ }\href
  {https://doi.org/10.1021/nl1021046} {\bibfield  {journal} {\bibinfo
  {journal} {Nano Lett.}\ }\textbf {\bibinfo {volume} {10}},\ \bibinfo {pages}
  {4067} (\bibinfo {year} {2010})}\BibitemShut {NoStop}%
\bibitem [{\citenamefont {Keerthi}\ \emph {et~al.}(2021)\citenamefont
  {Keerthi}, \citenamefont {Goutham}, \citenamefont {You}, \citenamefont
  {Iamprasertkun}, \citenamefont {Dryfe}, \citenamefont {Geim},\ and\
  \citenamefont {Radha}}]{Keerthi2021}%
  \BibitemOpen
  \bibfield  {author} {\bibinfo {author} {\bibfnamefont {A.}~\bibnamefont
  {Keerthi}}, \bibinfo {author} {\bibfnamefont {S.}~\bibnamefont {Goutham}},
  \bibinfo {author} {\bibfnamefont {Y.}~\bibnamefont {You}}, \bibinfo {author}
  {\bibfnamefont {P.}~\bibnamefont {Iamprasertkun}}, \bibinfo {author}
  {\bibfnamefont {R.~A.}\ \bibnamefont {Dryfe}}, \bibinfo {author}
  {\bibfnamefont {A.~K.}\ \bibnamefont {Geim}},\ and\ \bibinfo {author}
  {\bibfnamefont {B.}~\bibnamefont {Radha}},\ }\href
  {https://doi.org/10.1038/s41467-021-23325-3} {\bibfield  {journal} {\bibinfo
  {journal} {Nat. Commun.}\ }\textbf {\bibinfo {volume} {12}},\ \bibinfo
  {pages} {3092} (\bibinfo {year} {2021})}\BibitemShut {NoStop}%
\bibitem [{\citenamefont {Chen}\ \emph {et~al.}(2021)\citenamefont {Chen},
  \citenamefont {Li}, \citenamefont {Omori}, \citenamefont {Yamaguchi},
  \citenamefont {Ikuta},\ and\ \citenamefont {Takahashi}}]{Chen2021}%
  \BibitemOpen
  \bibfield  {author} {\bibinfo {author} {\bibfnamefont {K.-T.}\ \bibnamefont
  {Chen}}, \bibinfo {author} {\bibfnamefont {Q.-Y.}\ \bibnamefont {Li}},
  \bibinfo {author} {\bibfnamefont {T.}~\bibnamefont {Omori}}, \bibinfo
  {author} {\bibfnamefont {Y.}~\bibnamefont {Yamaguchi}}, \bibinfo {author}
  {\bibfnamefont {T.}~\bibnamefont {Ikuta}},\ and\ \bibinfo {author}
  {\bibfnamefont {K.}~\bibnamefont {Takahashi}},\ }\href
  {https://doi.org/10.1016/j.carbon.2021.12.048} {\bibfield  {journal}
  {\bibinfo  {journal} {Carbon}\ }\textbf {\bibinfo {volume} {189}},\ \bibinfo
  {pages} {162} (\bibinfo {year} {2021})}\BibitemShut {NoStop}%
\bibitem [{\citenamefont {Huang}\ \emph {et~al.}(2008)\citenamefont {Huang},
  \citenamefont {Cottin-Bizonne}, \citenamefont {Ybert},\ and\ \citenamefont
  {Bocquet}}]{Huang2008}%
  \BibitemOpen
  \bibfield  {author} {\bibinfo {author} {\bibfnamefont {D.~M.}\ \bibnamefont
  {Huang}}, \bibinfo {author} {\bibfnamefont {C.}~\bibnamefont
  {Cottin-Bizonne}}, \bibinfo {author} {\bibfnamefont {C.}~\bibnamefont
  {Ybert}},\ and\ \bibinfo {author} {\bibfnamefont {L.}~\bibnamefont
  {Bocquet}},\ }\href {https://doi.org/10.1021/la7021787} {\bibfield  {journal}
  {\bibinfo  {journal} {Langmuir}\ }\textbf {\bibinfo {volume} {24}},\ \bibinfo
  {pages} {1442} (\bibinfo {year} {2008})}\BibitemShut {NoStop}%
\bibitem [{\citenamefont {Ogawa}\ \emph {et~al.}(2019)\citenamefont {Ogawa},
  \citenamefont {Oga}, \citenamefont {Kusudo}, \citenamefont {Yamaguchi},
  \citenamefont {Omori}, \citenamefont {Merabia},\ and\ \citenamefont
  {Joly}}]{Ogawa2019}%
  \BibitemOpen
  \bibfield  {author} {\bibinfo {author} {\bibfnamefont {K.}~\bibnamefont
  {Ogawa}}, \bibinfo {author} {\bibfnamefont {H.}~\bibnamefont {Oga}}, \bibinfo
  {author} {\bibfnamefont {H.}~\bibnamefont {Kusudo}}, \bibinfo {author}
  {\bibfnamefont {Y.}~\bibnamefont {Yamaguchi}}, \bibinfo {author}
  {\bibfnamefont {T.}~\bibnamefont {Omori}}, \bibinfo {author} {\bibfnamefont
  {S.}~\bibnamefont {Merabia}},\ and\ \bibinfo {author} {\bibfnamefont
  {L.}~\bibnamefont {Joly}},\ }\href
  {https://doi.org/10.1103/PhysRevE.100.023101} {\bibfield  {journal} {\bibinfo
   {journal} {Phys. Rev. E}\ }\textbf {\bibinfo {volume} {100}},\ \bibinfo
  {pages} {023101} (\bibinfo {year} {2019})}\BibitemShut {NoStop}%
\bibitem [{\citenamefont {Thompson}\ and\ \citenamefont
  {Robbins}(1990)}]{Thompson1990}%
  \BibitemOpen
  \bibfield  {author} {\bibinfo {author} {\bibfnamefont {P.~A.}\ \bibnamefont
  {Thompson}}\ and\ \bibinfo {author} {\bibfnamefont {M.~O.}\ \bibnamefont
  {Robbins}},\ }\href {https://doi.org/10.1103/PhysRevA.41.6830} {\bibfield
  {journal} {\bibinfo  {journal} {Phys. Rev. A}\ }\textbf {\bibinfo {volume}
  {41}},\ \bibinfo {pages} {6830} (\bibinfo {year} {1990})}\BibitemShut
  {NoStop}%
\bibitem [{\citenamefont {Thompson}\ and\ \citenamefont
  {Troian}(1997)}]{Thompson1997}%
  \BibitemOpen
  \bibfield  {author} {\bibinfo {author} {\bibfnamefont {P.~A.}\ \bibnamefont
  {Thompson}}\ and\ \bibinfo {author} {\bibfnamefont {S.~M.}\ \bibnamefont
  {Troian}},\ }\href {https://doi.org/10.1038/38686} {\bibfield  {journal}
  {\bibinfo  {journal} {Nature}\ }\textbf {\bibinfo {volume} {389}},\ \bibinfo
  {pages} {360} (\bibinfo {year} {1997})}\BibitemShut {NoStop}%
\bibitem [{\citenamefont {Barrat}\ and\ \citenamefont
  {Bocquet}(1999)}]{Barrat1999}%
  \BibitemOpen
  \bibfield  {author} {\bibinfo {author} {\bibfnamefont {J.~L.}\ \bibnamefont
  {Barrat}}\ and\ \bibinfo {author} {\bibfnamefont {L.}~\bibnamefont
  {Bocquet}},\ }\href
  {https://pubs.rsc.org/en/content/articlelanding/1999/fd/a809733j} {\bibfield
  {journal} {\bibinfo  {journal} {Faraday Discuss.}\ }\textbf {\bibinfo
  {volume} {112}},\ \bibinfo {pages} {119} (\bibinfo {year}
  {1999})}\BibitemShut {NoStop}%
\bibitem [{\citenamefont {Cieplak}\ \emph {et~al.}(2001)\citenamefont
  {Cieplak}, \citenamefont {Koplik},\ and\ \citenamefont
  {Banavar}}]{Cieplak2001}%
  \BibitemOpen
  \bibfield  {author} {\bibinfo {author} {\bibfnamefont {M.}~\bibnamefont
  {Cieplak}}, \bibinfo {author} {\bibfnamefont {J.}~\bibnamefont {Koplik}},\
  and\ \bibinfo {author} {\bibfnamefont {J.~R.}\ \bibnamefont {Banavar}},\
  }\href {https://doi.org/10.1103/PhysRevLett.86.803} {\bibfield  {journal}
  {\bibinfo  {journal} {Phys. Rev. Lett.}\ }\textbf {\bibinfo {volume} {86}},\
  \bibinfo {pages} {803} (\bibinfo {year} {2001})}\BibitemShut {NoStop}%
\bibitem [{\citenamefont {Kannam}\ \emph {et~al.}(2013)\citenamefont {Kannam},
  \citenamefont {Todd}, \citenamefont {Hansen},\ and\ \citenamefont
  {Daivis}}]{Kannam2013}%
  \BibitemOpen
  \bibfield  {author} {\bibinfo {author} {\bibfnamefont {S.~K.}\ \bibnamefont
  {Kannam}}, \bibinfo {author} {\bibfnamefont {B.~D.}\ \bibnamefont {Todd}},
  \bibinfo {author} {\bibfnamefont {J.~S.}\ \bibnamefont {Hansen}},\ and\
  \bibinfo {author} {\bibfnamefont {P.~J.}\ \bibnamefont {Daivis}},\ }\href
  {https://doi.org/10.1063/1.4793396} {\bibfield  {journal} {\bibinfo
  {journal} {J. Chem. Phys.}\ }\textbf {\bibinfo {volume} {138}},\ \bibinfo
  {pages} {094701} (\bibinfo {year} {2013})}\BibitemShut {NoStop}%
\bibitem [{\citenamefont {Bhatia}\ and\ \citenamefont
  {Nicholson}(2013)}]{Bhatia2013}%
  \BibitemOpen
  \bibfield  {author} {\bibinfo {author} {\bibfnamefont {S.~K.}\ \bibnamefont
  {Bhatia}}\ and\ \bibinfo {author} {\bibfnamefont {D.}~\bibnamefont
  {Nicholson}},\ }\href {https://pubs.acs.org/doi/10.1021/la403445j} {\bibfield
   {journal} {\bibinfo  {journal} {Langmuir}\ }\textbf {\bibinfo {volume}
  {29}},\ \bibinfo {pages} {14519} (\bibinfo {year} {2013})}\BibitemShut
  {NoStop}%
\bibitem [{\citenamefont {Tocci}\ \emph {et~al.}(2014)\citenamefont {Tocci},
  \citenamefont {Joly},\ and\ \citenamefont {Michaelides}}]{Tocci2014}%
  \BibitemOpen
  \bibfield  {author} {\bibinfo {author} {\bibfnamefont {G.}~\bibnamefont
  {Tocci}}, \bibinfo {author} {\bibfnamefont {L.}~\bibnamefont {Joly}},\ and\
  \bibinfo {author} {\bibfnamefont {A.}~\bibnamefont {Michaelides}},\ }\href
  {https://doi.org/10.1021/nl502837d} {\bibfield  {journal} {\bibinfo
  {journal} {Nano Lett.}\ }\textbf {\bibinfo {volume} {14}},\ \bibinfo {pages}
  {6872} (\bibinfo {year} {2014})}\BibitemShut {NoStop}%
\bibitem [{\citenamefont {Guo}\ \emph {et~al.}(2016)\citenamefont {Guo},
  \citenamefont {Chen},\ and\ \citenamefont {Robbins}}]{Guo2016}%
  \BibitemOpen
  \bibfield  {author} {\bibinfo {author} {\bibfnamefont {L.}~\bibnamefont
  {Guo}}, \bibinfo {author} {\bibfnamefont {S.}~\bibnamefont {Chen}},\ and\
  \bibinfo {author} {\bibfnamefont {M.~O.}\ \bibnamefont {Robbins}},\ }\href
  {https://doi.org/10.1140/epjst/e2016-60146-3} {\bibfield  {journal} {\bibinfo
   {journal} {Eur. Phys. J. Spec. Top.}\ }\textbf {\bibinfo {volume} {225}},\
  \bibinfo {pages} {1551} (\bibinfo {year} {2016})}\BibitemShut {NoStop}%
\bibitem [{\citenamefont {Nakaoka}\ \emph
  {et~al.}(2017{\natexlab{a}})\citenamefont {Nakaoka}, \citenamefont
  {Yamaguchi}, \citenamefont {Omori},\ and\ \citenamefont
  {Joly}}]{Nakaoka2017}%
  \BibitemOpen
  \bibfield  {author} {\bibinfo {author} {\bibfnamefont {S.}~\bibnamefont
  {Nakaoka}}, \bibinfo {author} {\bibfnamefont {Y.}~\bibnamefont {Yamaguchi}},
  \bibinfo {author} {\bibfnamefont {T.}~\bibnamefont {Omori}},\ and\ \bibinfo
  {author} {\bibfnamefont {L.}~\bibnamefont {Joly}},\ }\href
  {https://doi.org/10.1063/1.4982617} {\bibfield  {journal} {\bibinfo
  {journal} {J. Chem. Phys.}\ }\textbf {\bibinfo {volume} {146}},\ \bibinfo
  {pages} {174702} (\bibinfo {year} {2017}{\natexlab{a}})}\BibitemShut
  {NoStop}%
\bibitem [{\citenamefont {Ewen}\ \emph {et~al.}(2019)\citenamefont {Ewen},
  \citenamefont {Gao}, \citenamefont {M{\"{u}}ser},\ and\ \citenamefont
  {Dini}}]{Ewen2019}%
  \BibitemOpen
  \bibfield  {author} {\bibinfo {author} {\bibfnamefont {J.~P.}\ \bibnamefont
  {Ewen}}, \bibinfo {author} {\bibfnamefont {H.}~\bibnamefont {Gao}}, \bibinfo
  {author} {\bibfnamefont {M.~H.}\ \bibnamefont {M{\"{u}}ser}},\ and\ \bibinfo
  {author} {\bibfnamefont {D.}~\bibnamefont {Dini}},\ }\href
  {https://doi.org/10.1039/c8cp07436d} {\bibfield  {journal} {\bibinfo
  {journal} {Phys. Chem. Chem. Phys.}\ }\textbf {\bibinfo {volume} {21}},\
  \bibinfo {pages} {5813} (\bibinfo {year} {2019})}\BibitemShut {NoStop}%
\bibitem [{\citenamefont {Kannam}\ \emph {et~al.}(2011)\citenamefont {Kannam},
  \citenamefont {Todd}, \citenamefont {Hansen},\ and\ \citenamefont
  {Daivis}}]{Kannam2011}%
  \BibitemOpen
  \bibfield  {author} {\bibinfo {author} {\bibfnamefont {S.~K.}\ \bibnamefont
  {Kannam}}, \bibinfo {author} {\bibfnamefont {B.~D.}\ \bibnamefont {Todd}},
  \bibinfo {author} {\bibfnamefont {J.~S.}\ \bibnamefont {Hansen}},\ and\
  \bibinfo {author} {\bibfnamefont {P.~J.}\ \bibnamefont {Daivis}},\ }\href
  {https://doi.org/10.1063/1.3648049} {\bibfield  {journal} {\bibinfo
  {journal} {J. Chem. Phys.}\ }\textbf {\bibinfo {volume} {135}},\ \bibinfo
  {pages} {144701} (\bibinfo {year} {2011})}\BibitemShut {NoStop}%
\bibitem [{\citenamefont {{Kumar Kannam}}\ \emph {et~al.}(2012)\citenamefont
  {{Kumar Kannam}}, \citenamefont {Todd}, \citenamefont {Hansen},\ and\
  \citenamefont {Daivis}}]{KumarKannam2012}%
  \BibitemOpen
  \bibfield  {author} {\bibinfo {author} {\bibfnamefont {S.}~\bibnamefont
  {{Kumar Kannam}}}, \bibinfo {author} {\bibfnamefont {B.~D.}\ \bibnamefont
  {Todd}}, \bibinfo {author} {\bibfnamefont {J.~S.}\ \bibnamefont {Hansen}},\
  and\ \bibinfo {author} {\bibfnamefont {P.~J.}\ \bibnamefont {Daivis}},\
  }\href {https://doi.org/10.1063/1.3675904} {\bibfield  {journal} {\bibinfo
  {journal} {J. Chem. Phys.}\ }\textbf {\bibinfo {volume} {136}},\ \bibinfo
  {pages} {024705} (\bibinfo {year} {2012})}\BibitemShut {NoStop}%
\bibitem [{\citenamefont {Herrero}\ \emph {et~al.}(2019)\citenamefont
  {Herrero}, \citenamefont {Omori}, \citenamefont {Yamaguchi},\ and\
  \citenamefont {Joly}}]{Herrero2019}%
  \BibitemOpen
  \bibfield  {author} {\bibinfo {author} {\bibfnamefont {C.}~\bibnamefont
  {Herrero}}, \bibinfo {author} {\bibfnamefont {T.}~\bibnamefont {Omori}},
  \bibinfo {author} {\bibfnamefont {Y.}~\bibnamefont {Yamaguchi}},\ and\
  \bibinfo {author} {\bibfnamefont {L.}~\bibnamefont {Joly}},\ }\href
  {https://doi.org/10.1063/1.5111966} {\bibfield  {journal} {\bibinfo
  {journal} {J. Chem. Phys.}\ }\textbf {\bibinfo {volume} {151}},\ \bibinfo
  {pages} {041103} (\bibinfo {year} {2019})}\BibitemShut {NoStop}%
\bibitem [{\citenamefont {Bocquet}\ and\ \citenamefont
  {Barrat}(1994)}]{Bocquet1994}%
  \BibitemOpen
  \bibfield  {author} {\bibinfo {author} {\bibfnamefont {L.}~\bibnamefont
  {Bocquet}}\ and\ \bibinfo {author} {\bibfnamefont {J.~L.}\ \bibnamefont
  {Barrat}},\ }\href {https://doi.org/10.1103/PhysRevE.49.3079} {\bibfield
  {journal} {\bibinfo  {journal} {Phys. Rev. E}\ }\textbf {\bibinfo {volume}
  {49}},\ \bibinfo {pages} {3079} (\bibinfo {year} {1994})}\BibitemShut
  {NoStop}%
\bibitem [{\citenamefont {Petravic}\ and\ \citenamefont
  {Harrowell}(2007)}]{Petravic2007}%
  \BibitemOpen
  \bibfield  {author} {\bibinfo {author} {\bibfnamefont {J.}~\bibnamefont
  {Petravic}}\ and\ \bibinfo {author} {\bibfnamefont {P.}~\bibnamefont
  {Harrowell}},\ }\href {https://doi.org/10.1063/1.2799186} {\bibfield
  {journal} {\bibinfo  {journal} {J. Chem. Phys.}\ }\textbf {\bibinfo {volume}
  {127}},\ \bibinfo {pages} {174706} (\bibinfo {year} {2007})}\BibitemShut
  {NoStop}%
\bibitem [{\citenamefont {Hansen}\ \emph {et~al.}(2011)\citenamefont {Hansen},
  \citenamefont {Todd},\ and\ \citenamefont {Daivis}}]{Hansen2011}%
  \BibitemOpen
  \bibfield  {author} {\bibinfo {author} {\bibfnamefont {J.~S.}\ \bibnamefont
  {Hansen}}, \bibinfo {author} {\bibfnamefont {B.~D.}\ \bibnamefont {Todd}},\
  and\ \bibinfo {author} {\bibfnamefont {P.~J.}\ \bibnamefont {Daivis}},\
  }\href {https://doi.org/10.1103/PhysRevE.84.016313} {\bibfield  {journal}
  {\bibinfo  {journal} {Phys. Rev. E}\ }\textbf {\bibinfo {volume} {84}},\
  \bibinfo {pages} {016313} (\bibinfo {year} {2011})}\BibitemShut {NoStop}%
\bibitem [{\citenamefont {Bocquet}\ and\ \citenamefont
  {Barrat}(2013)}]{Bocquet2013}%
  \BibitemOpen
  \bibfield  {author} {\bibinfo {author} {\bibfnamefont {L.}~\bibnamefont
  {Bocquet}}\ and\ \bibinfo {author} {\bibfnamefont {J.~L.}\ \bibnamefont
  {Barrat}},\ }\href {https://doi.org/10.1063/1.4816006} {\bibfield  {journal}
  {\bibinfo  {journal} {J. Chem. Phys.}\ }\textbf {\bibinfo {volume} {139}},\
  \bibinfo {pages} {044704} (\bibinfo {year} {2013})}\BibitemShut {NoStop}%
\bibitem [{\citenamefont {Huang}\ and\ \citenamefont
  {Szlufarska}(2014)}]{Huang2014}%
  \BibitemOpen
  \bibfield  {author} {\bibinfo {author} {\bibfnamefont {K.}~\bibnamefont
  {Huang}}\ and\ \bibinfo {author} {\bibfnamefont {I.}~\bibnamefont
  {Szlufarska}},\ }\href {https://doi.org/10.1103/PhysRevE.89.032119}
  {\bibfield  {journal} {\bibinfo  {journal} {Phys. Rev. E}\ }\textbf {\bibinfo
  {volume} {89}},\ \bibinfo {pages} {032119} (\bibinfo {year}
  {2014})}\BibitemShut {NoStop}%
\bibitem [{\citenamefont {Sam}\ \emph {et~al.}(2018)\citenamefont {Sam},
  \citenamefont {Hartkamp}, \citenamefont {Kannam},\ and\ \citenamefont
  {Sathian}}]{Sam2018}%
  \BibitemOpen
  \bibfield  {author} {\bibinfo {author} {\bibfnamefont {A.}~\bibnamefont
  {Sam}}, \bibinfo {author} {\bibfnamefont {R.}~\bibnamefont {Hartkamp}},
  \bibinfo {author} {\bibfnamefont {S.~K.}\ \bibnamefont {Kannam}},\ and\
  \bibinfo {author} {\bibfnamefont {S.~P.}\ \bibnamefont {Sathian}},\ }\href
  {https://doi.org/10.1088/1361-6528/aae0bd} {\bibfield  {journal} {\bibinfo
  {journal} {Nanotechnology}\ }\textbf {\bibinfo {volume} {29}},\ \bibinfo
  {pages} {485404} (\bibinfo {year} {2018})}\BibitemShut {NoStop}%
\bibitem [{\citenamefont {Oga}\ \emph {et~al.}(2019)\citenamefont {Oga},
  \citenamefont {Yamaguchi}, \citenamefont {Omori}, \citenamefont {Merabia},\
  and\ \citenamefont {Joly}}]{Oga2019}%
  \BibitemOpen
  \bibfield  {author} {\bibinfo {author} {\bibfnamefont {H.}~\bibnamefont
  {Oga}}, \bibinfo {author} {\bibfnamefont {Y.}~\bibnamefont {Yamaguchi}},
  \bibinfo {author} {\bibfnamefont {T.}~\bibnamefont {Omori}}, \bibinfo
  {author} {\bibfnamefont {S.}~\bibnamefont {Merabia}},\ and\ \bibinfo {author}
  {\bibfnamefont {L.}~\bibnamefont {Joly}},\ }\href
  {https://doi.org/10.1063/1.5104335} {\bibfield  {journal} {\bibinfo
  {journal} {J. Chem. Phys.}\ }\textbf {\bibinfo {volume} {151}},\ \bibinfo
  {pages} {054502} (\bibinfo {year} {2019})}\BibitemShut {NoStop}%
\bibitem [{\citenamefont {Varghese}\ \emph {et~al.}(2021)\citenamefont
  {Varghese}, \citenamefont {Hansen},\ and\ \citenamefont
  {Todd}}]{Varghese2021}%
  \BibitemOpen
  \bibfield  {author} {\bibinfo {author} {\bibfnamefont {S.}~\bibnamefont
  {Varghese}}, \bibinfo {author} {\bibfnamefont {J.~S.}\ \bibnamefont
  {Hansen}},\ and\ \bibinfo {author} {\bibfnamefont {B.~D.}\ \bibnamefont
  {Todd}},\ }\href {https://doi.org/10.1063/5.0040191} {\bibfield  {journal}
  {\bibinfo  {journal} {J. Chem. Phys.}\ }\textbf {\bibinfo {volume} {154}},\
  \bibinfo {pages} {184707} (\bibinfo {year} {2021})}\BibitemShut {NoStop}%
\bibitem [{\citenamefont {Nakano}\ and\ \citenamefont
  {Sasa}(2020)}]{Nakano2020}%
  \BibitemOpen
  \bibfield  {author} {\bibinfo {author} {\bibfnamefont {H.}~\bibnamefont
  {Nakano}}\ and\ \bibinfo {author} {\bibfnamefont {S.}~\bibnamefont {Sasa}},\
  }\href {https://doi.org/10.1103/PhysRevE.101.033109} {\bibfield  {journal}
  {\bibinfo  {journal} {Phys. Rev. E}\ }\textbf {\bibinfo {volume} {101}},\
  \bibinfo {pages} {033109} (\bibinfo {year} {2020})}\BibitemShut {NoStop}%
\bibitem [{\citenamefont {Sokhan}\ and\ \citenamefont
  {Quirke}(2008)}]{Sokhan2008}%
  \BibitemOpen
  \bibfield  {author} {\bibinfo {author} {\bibfnamefont {V.~P.}\ \bibnamefont
  {Sokhan}}\ and\ \bibinfo {author} {\bibfnamefont {N.}~\bibnamefont
  {Quirke}},\ }\href {https://doi.org/10.1103/PhysRevE.78.015301} {\bibfield
  {journal} {\bibinfo  {journal} {Phys. Rev. E}\ }\textbf {\bibinfo {volume}
  {78}},\ \bibinfo {pages} {015301(R)} (\bibinfo {year} {2008})}\BibitemShut
  {NoStop}%
\bibitem [{\citenamefont {Hadjiconstantinou}\ and\ \citenamefont
  {Swisher}(2022)}]{Hadjiconstantinou2022}%
  \BibitemOpen
  \bibfield  {author} {\bibinfo {author} {\bibfnamefont {N.~G.}\ \bibnamefont
  {Hadjiconstantinou}}\ and\ \bibinfo {author} {\bibfnamefont {M.~M.}\
  \bibnamefont {Swisher}},\ }\href
  {https://doi.org/10.1103/PhysRevFluids.7.114203} {\bibfield  {journal}
  {\bibinfo  {journal} {Phys. Rev. Fluids}\ }\textbf {\bibinfo {volume} {7}},\
  \bibinfo {pages} {114203} (\bibinfo {year} {2022})}\BibitemShut {NoStop}%
\bibitem [{\citenamefont {Evans}\ and\ \citenamefont
  {Morriss}(2008)}]{Evans2008book}%
  \BibitemOpen
  \bibfield  {author} {\bibinfo {author} {\bibfnamefont {D.}~\bibnamefont
  {Evans}}\ and\ \bibinfo {author} {\bibfnamefont {G.}~\bibnamefont
  {Morriss}},\ }\href
  {https://doi.org/http://dx.doi.org/10.1017/CBO9780511535307} {\emph {\bibinfo
  {title} {Statistical Mechanics of Nonequilibrium Liquids}}},\ \bibinfo
  {edition} {2nd}\ ed.\ (\bibinfo  {publisher} {Cambridge University Press},\
  \bibinfo {year} {2008})\ pp.\ \bibinfo {pages} {71--72}\BibitemShut {NoStop}%
\bibitem [{\citenamefont {Espa{\~{n}}ol}\ \emph {et~al.}(2019)\citenamefont
  {Espa{\~{n}}ol}, \citenamefont {{De La Torre}},\ and\ \citenamefont
  {Duque-Zumajo}}]{Espanol2019}%
  \BibitemOpen
  \bibfield  {author} {\bibinfo {author} {\bibfnamefont {P.}~\bibnamefont
  {Espa{\~{n}}ol}}, \bibinfo {author} {\bibfnamefont {J.~A.}\ \bibnamefont {{De
  La Torre}}},\ and\ \bibinfo {author} {\bibfnamefont {D.}~\bibnamefont
  {Duque-Zumajo}},\ }\href {https://doi.org/10.1103/PhysRevE.99.022126}
  {\bibfield  {journal} {\bibinfo  {journal} {Phys. Rev. E}\ }\textbf {\bibinfo
  {volume} {99}},\ \bibinfo {pages} {022126} (\bibinfo {year}
  {2019})}\BibitemShut {NoStop}%
\bibitem [{\citenamefont {Merabia}\ and\ \citenamefont
  {Termentzidis}(2012)}]{Merabia2012}%
  \BibitemOpen
  \bibfield  {author} {\bibinfo {author} {\bibfnamefont {S.}~\bibnamefont
  {Merabia}}\ and\ \bibinfo {author} {\bibfnamefont {K.}~\bibnamefont
  {Termentzidis}},\ }\href {https://doi.org/10.1103/PhysRevB.86.094303}
  {\bibfield  {journal} {\bibinfo  {journal} {Phys. Rev. B}\ }\textbf {\bibinfo
  {volume} {86}},\ \bibinfo {pages} {094303} (\bibinfo {year}
  {2012})}\BibitemShut {NoStop}%
\bibitem [{\citenamefont {Nakaoka}\ \emph
  {et~al.}(2017{\natexlab{b}})\citenamefont {Nakaoka}, \citenamefont
  {Yamaguchi}, \citenamefont {Omori},\ and\ \citenamefont
  {Joly}}]{Nakaoka2017b}%
  \BibitemOpen
  \bibfield  {author} {\bibinfo {author} {\bibfnamefont {S.}~\bibnamefont
  {Nakaoka}}, \bibinfo {author} {\bibfnamefont {Y.}~\bibnamefont {Yamaguchi}},
  \bibinfo {author} {\bibfnamefont {T.}~\bibnamefont {Omori}},\ and\ \bibinfo
  {author} {\bibfnamefont {L.}~\bibnamefont {Joly}},\ }\href
  {https://doi.org/10.1299/mel.17-00422} {\bibfield  {journal} {\bibinfo
  {journal} {Mech. Eng. Lett.}\ }\textbf {\bibinfo {volume} {3}},\ \bibinfo
  {pages} {17} (\bibinfo {year} {2017}{\natexlab{b}})}\BibitemShut {NoStop}%
\bibitem [{\citenamefont {Schulz}\ \emph {et~al.}(2020)\citenamefont {Schulz},
  \citenamefont {Schlaich}, \citenamefont {Heyden}, \citenamefont {Netz},\ and\
  \citenamefont {Kappler}}]{Schulz2020}%
  \BibitemOpen
  \bibfield  {author} {\bibinfo {author} {\bibfnamefont {J.~C.~F.}\
  \bibnamefont {Schulz}}, \bibinfo {author} {\bibfnamefont {A.}~\bibnamefont
  {Schlaich}}, \bibinfo {author} {\bibfnamefont {M.}~\bibnamefont {Heyden}},
  \bibinfo {author} {\bibfnamefont {R.~R.}\ \bibnamefont {Netz}},\ and\
  \bibinfo {author} {\bibfnamefont {J.}~\bibnamefont {Kappler}},\ }\href
  {https://doi.org/10.1103/PhysRevFluids.5.103301} {\bibfield  {journal}
  {\bibinfo  {journal} {Phys. Rev. Fluids}\ }\textbf {\bibinfo {volume} {5}},\
  \bibinfo {pages} {103301} (\bibinfo {year} {2020})}\BibitemShut {NoStop}%
\bibitem [{\citenamefont {Omori}\ \emph {et~al.}(2019)\citenamefont {Omori},
  \citenamefont {Inoue}, \citenamefont {Joly}, \citenamefont {Merabia},\ and\
  \citenamefont {Yamaguchi}}]{Omori2019}%
  \BibitemOpen
  \bibfield  {author} {\bibinfo {author} {\bibfnamefont {T.}~\bibnamefont
  {Omori}}, \bibinfo {author} {\bibfnamefont {N.}~\bibnamefont {Inoue}},
  \bibinfo {author} {\bibfnamefont {L.}~\bibnamefont {Joly}}, \bibinfo {author}
  {\bibfnamefont {S.}~\bibnamefont {Merabia}},\ and\ \bibinfo {author}
  {\bibfnamefont {Y.}~\bibnamefont {Yamaguchi}},\ }\href
  {https://doi.org/10.1103/PhysRevFluids.4.114201} {\bibfield  {journal}
  {\bibinfo  {journal} {Phys. Rev. Fluids}\ }\textbf {\bibinfo {volume} {4}},\
  \bibinfo {pages} {114201} (\bibinfo {year} {2019})}\BibitemShut {NoStop}%
\bibitem [{\citenamefont {Oga}\ \emph {et~al.}(2021)\citenamefont {Oga},
  \citenamefont {Omori}, \citenamefont {Herrero}, \citenamefont {Merabia},
  \citenamefont {Joly},\ and\ \citenamefont {Yamaguchi}}]{Oga2021}%
  \BibitemOpen
  \bibfield  {author} {\bibinfo {author} {\bibfnamefont {H.}~\bibnamefont
  {Oga}}, \bibinfo {author} {\bibfnamefont {T.}~\bibnamefont {Omori}}, \bibinfo
  {author} {\bibfnamefont {C.}~\bibnamefont {Herrero}}, \bibinfo {author}
  {\bibfnamefont {S.}~\bibnamefont {Merabia}}, \bibinfo {author} {\bibfnamefont
  {L.}~\bibnamefont {Joly}},\ and\ \bibinfo {author} {\bibfnamefont
  {Y.}~\bibnamefont {Yamaguchi}},\ }\href
  {https://doi.org/10.1103/PhysRevResearch.3.L032019} {\bibfield  {journal}
  {\bibinfo  {journal} {Phys. Rev. Res.}\ }\textbf {\bibinfo {volume} {3}},\
  \bibinfo {pages} {L032019} (\bibinfo {year} {2021})}\BibitemShut {NoStop}%
\bibitem [{\citenamefont {Nakano}\ and\ \citenamefont
  {Sasa}(2019{\natexlab{a}})}]{Nakano2019}%
  \BibitemOpen
  \bibfield  {author} {\bibinfo {author} {\bibfnamefont {H.}~\bibnamefont
  {Nakano}}\ and\ \bibinfo {author} {\bibfnamefont {S.}~\bibnamefont {Sasa}},\
  }\href {https://doi.org/10.1007/s10955-019-02302-7} {\bibfield  {journal}
  {\bibinfo  {journal} {J. Stat. Phys.}\ }\textbf {\bibinfo {volume} {176}},\
  \bibinfo {pages} {312} (\bibinfo {year} {2019}{\natexlab{a}})}\BibitemShut
  {NoStop}%
\bibitem [{\citenamefont {Nakano}\ and\ \citenamefont
  {Sasa}(2019{\natexlab{b}})}]{Nakano2019a}%
  \BibitemOpen
  \bibfield  {author} {\bibinfo {author} {\bibfnamefont {H.}~\bibnamefont
  {Nakano}}\ and\ \bibinfo {author} {\bibfnamefont {S.}~\bibnamefont {Sasa}},\
  }\href {https://doi.org/10.1103/PhysRevE.99.013106} {\bibfield  {journal}
  {\bibinfo  {journal} {Phys. Rev. E}\ }\textbf {\bibinfo {volume} {99}},\
  \bibinfo {pages} {013106} (\bibinfo {year} {2019}{\natexlab{b}})}\BibitemShut
  {NoStop}%
\bibitem [{\citenamefont {Thompson}\ \emph {et~al.}(2022)\citenamefont
  {Thompson}, \citenamefont {Aktulga}, \citenamefont {Berger}, \citenamefont
  {Bolintineanu}, \citenamefont {Brown}, \citenamefont {Crozier}, \citenamefont
  {{in't Veld}}, \citenamefont {Kohlmeyer}, \citenamefont {Moore},
  \citenamefont {Nguyen}, \citenamefont {Shan}, \citenamefont {Stevens},
  \citenamefont {Tranchida}, \citenamefont {Trott},\ and\ \citenamefont
  {Plimpton}}]{THOMPSON2022108171}%
  \BibitemOpen
  \bibfield  {author} {\bibinfo {author} {\bibfnamefont {A.~P.}\ \bibnamefont
  {Thompson}}, \bibinfo {author} {\bibfnamefont {H.~M.}\ \bibnamefont
  {Aktulga}}, \bibinfo {author} {\bibfnamefont {R.}~\bibnamefont {Berger}},
  \bibinfo {author} {\bibfnamefont {D.~S.}\ \bibnamefont {Bolintineanu}},
  \bibinfo {author} {\bibfnamefont {W.~M.}\ \bibnamefont {Brown}}, \bibinfo
  {author} {\bibfnamefont {P.~S.}\ \bibnamefont {Crozier}}, \bibinfo {author}
  {\bibfnamefont {P.~J.}\ \bibnamefont {{in't Veld}}}, \bibinfo {author}
  {\bibfnamefont {A.}~\bibnamefont {Kohlmeyer}}, \bibinfo {author}
  {\bibfnamefont {S.~G.}\ \bibnamefont {Moore}}, \bibinfo {author}
  {\bibfnamefont {T.~D.}\ \bibnamefont {Nguyen}}, \bibinfo {author}
  {\bibfnamefont {R.}~\bibnamefont {Shan}}, \bibinfo {author} {\bibfnamefont
  {M.~J.}\ \bibnamefont {Stevens}}, \bibinfo {author} {\bibfnamefont
  {J.}~\bibnamefont {Tranchida}}, \bibinfo {author} {\bibfnamefont
  {C.}~\bibnamefont {Trott}},\ and\ \bibinfo {author} {\bibfnamefont {S.~J.}\
  \bibnamefont {Plimpton}},\ }\href
  {https://doi.org/https://doi.org/10.1016/j.cpc.2021.108171} {\bibfield
  {journal} {\bibinfo  {journal} {Comput. Phys. Commun.}\ }\textbf {\bibinfo
  {volume} {271}},\ \bibinfo {pages} {108171} (\bibinfo {year}
  {2022})}\BibitemShut {NoStop}%
\bibitem [{\citenamefont {Nishida}\ \emph {et~al.}(2014)\citenamefont
  {Nishida}, \citenamefont {Surblys}, \citenamefont {Yamaguchi}, \citenamefont
  {Kuroda}, \citenamefont {Kagawa}, \citenamefont {Nakajima},\ and\
  \citenamefont {Fujimura}}]{Nishida2014}%
  \BibitemOpen
  \bibfield  {author} {\bibinfo {author} {\bibfnamefont {S.}~\bibnamefont
  {Nishida}}, \bibinfo {author} {\bibfnamefont {D.}~\bibnamefont {Surblys}},
  \bibinfo {author} {\bibfnamefont {Y.}~\bibnamefont {Yamaguchi}}, \bibinfo
  {author} {\bibfnamefont {K.}~\bibnamefont {Kuroda}}, \bibinfo {author}
  {\bibfnamefont {M.}~\bibnamefont {Kagawa}}, \bibinfo {author} {\bibfnamefont
  {T.}~\bibnamefont {Nakajima}},\ and\ \bibinfo {author} {\bibfnamefont
  {H.}~\bibnamefont {Fujimura}},\ }\href {https://doi.org/10.1063/1.4865254}
  {\bibfield  {journal} {\bibinfo  {journal} {J. Chem. Phys.}\ }\textbf
  {\bibinfo {volume} {140}},\ \bibinfo {pages} {074707} (\bibinfo {year}
  {2014})}\BibitemShut {NoStop}%
\bibitem [{\citenamefont {Nakaoka}\ \emph {et~al.}(2015)\citenamefont
  {Nakaoka}, \citenamefont {Yamaguchi}, \citenamefont {Omori}, \citenamefont
  {Kagawa}, \citenamefont {Nakajima},\ and\ \citenamefont
  {Fujimura}}]{Nakaoka2015}%
  \BibitemOpen
  \bibfield  {author} {\bibinfo {author} {\bibfnamefont {S.}~\bibnamefont
  {Nakaoka}}, \bibinfo {author} {\bibfnamefont {Y.}~\bibnamefont {Yamaguchi}},
  \bibinfo {author} {\bibfnamefont {T.}~\bibnamefont {Omori}}, \bibinfo
  {author} {\bibfnamefont {M.}~\bibnamefont {Kagawa}}, \bibinfo {author}
  {\bibfnamefont {T.}~\bibnamefont {Nakajima}},\ and\ \bibinfo {author}
  {\bibfnamefont {H.}~\bibnamefont {Fujimura}},\ }\href
  {https://doi.org/10.1103/PhysRevE.92.022402} {\bibfield  {journal} {\bibinfo
  {journal} {Phys. Rev. E}\ }\textbf {\bibinfo {volume} {92}},\ \bibinfo
  {pages} {022402} (\bibinfo {year} {2015})}\BibitemShut {NoStop}%
\end{thebibliography}
\end{document}